\documentclass[%
 pre,
 amsmath,amssymb,
 reprint,%
]{revtex4-1}

\usepackage{amsmath}
\usepackage{graphicx}
\usepackage{dcolumn}
\usepackage{bm}
\newcommand\Rh{{\text{Rh}}}
\newcommand\Rey{{\text{Re}}}
\usepackage{caption}
\usepackage{subcaption}
\usepackage{epstopdf} 
\usepackage{graphics}
\usepackage{xcolor}
\usepackage{tikz}
\usetikzlibrary{shapes.geometric}
\DeclareRobustCommand\mytikzdot{\tikz\protect\draw[black,fill=black] (0,0) circle (.5ex); }
\DeclareMathOperator\erf{erf}

\def\Plus{\texttt{+}}

\begin{document}


\title[Spatial extreme values of vorticity and velocity gradients in two-dimensional turbulent flows]{Spatial extreme values of vorticity and velocity gradients in two-dimensional turbulent flows}

\author{Kannabiran Seshasayanan}
\affiliation{ 
Department of Physics, IIT Kharagpur, Kharagpur, West Bengal, India
}%

\date{\today}

\begin{abstract}
We study the distribution of spatial extrema of vorticity and the determinant of the strain rate tensor for a two-dimensional turbulent flow forced by a Kolmogorov forcing. 
The distribution of these quantities follow non-Gaussian behaviour and they do not fall into the Generalised Extreme value distributions. It is found that for the truncated Euler equations the spatial extrema of vorticity and strain rate tensor are well described by the Gumbel distribution. The spatial extrema for the vorticity is found to be at the core of the vortices while the velocity gradients are found near the edges of the vortices or at the shear layers in the regions between the vortices. Temporal correlations of the velocity gradients shed light on the extreme value distributions obtained for turbulence and the truncated Euler equations. 
\end{abstract}

\maketitle

\section{\label{sec:level1} Introduction}

Turbulent flows display fluctuations across many temporal and spatial scales. Fluctuations that are many times the standard deviation of the signal are termed as extreme events and they have been studied widely in a variety of systems. Three-dimensional turbulence exhibits large fluctuations in the local vorticity due to the formation of thin filaments. The vorticity of such filaments are orders of magnitude larger than the rms values and these are spatial localised structures where dissipation effects are strong \cite{yeung2015extreme, buaria2019extreme, yeung2020advancing}. 
Characterising such events is important for the study of small scale intermittency properties of turbulent flows. Statistics of extreme events have also been studied in systems such as, climate in the form of extreme temperature events \cite{easterling2000climate, ragone2018computation}, study of rogue and freak waves in oceanography \cite{dysthe2008oceanic, onorato2013rogue}, power fluctuations in connected grids \cite{grid_fluctuations}, polymer stretching \cite{picardo2023polymers} etc. Mitigation of rare events where the focus is to predict and control these events have also been studied, for example in \cite{farazmand2019closed, sapsis2021statistics} for turbulent flows. 

 Extreme values of a series of identical independent random numbers fall into one of the three general classes of distributions, Gumbel, Fr{\'e}chet and Weibull also known as the Fisher-Tippett-Gnedenko (FTG) theorem. The distribution of the random number decides which of the class of extreme value distribution the extremes fall into. If the random numbers are weakly correlated, in some situations it can also fall into the general class of extreme value distributions \cite{antal20011,majumdar2020extreme}, whereas for strongly correlated systems no such general result exists. For spatio-temporal systems, the study of spatial extreme values have been studied for example in KPZ, Edwards-Wilkinson model \cite{raychaudhuri2001maximal, gyorgyi2003statistics, rambeau2010extremal}. In certain systems where the limiting distribution of the spatial extremes could be obtained analytically, it was found that they do not fall under the known extreme value distributions of the FTG theorem, \cite{majumdar2005airy}. 


Recently it has been shown that extreme values of velocity gradients are linked to the transitions between different turbulent flows, \cite{seshasayanan2020onset}. Turbulent flows can display transitions between different states as an external parameter is varied, see \cite{alexakis2018cascades} for a detailed review. 
The system transitions from one turbulent state to another by the growth of perturbations on a turbulent background. This leads to a non-monotonic growth of the perturbations leading to intermittency in the growth rate. In rapidly rotating flows, it was shown that the onset of three-dimensionality in certain regimes is controlled by the minimum of the vorticity field while in other regimes, large values of the strain rate play an important role in this transition \cite{seshasayanan2020onset}. Such extreme fluctuations can govern the transitions between 2D-3D flows in thin-layer turbulence and in other turbulent transitions, \cite{seshasayanan2014edge,benavides2017critical,alexakis2018cascades}. The intermittency properties of the growth rate of the perturbations will be governed by the statistics of the underlying flow structures, as seen in recent model studies \cite{van2021intermittency, van20231}.  
 In turbulent flows where sub-critical branches, multi-branch solutions are found, large amplitude perturbations in vorticity or shear could decide the solution branch chosen by the system, \cite{alexakis2015rotating, favier2019subcritical}. In confined flows with large scale condensation, the existence of strong fluctuations in vorticity or shear could possibly trigger reversals and such systems also show $1/f$-type statistics, \cite{herault20151, dallas2020transitions}. In order to understand the distribution of such extreme fluctuations, we concentrate upon the case of two-dimensional turbulence, and more specifically we study the properties of distribution of the spatial extremes of vorticity and shear. 

In many models of two-dimensional turbulence a large scale friction parameter is introduced to model the confinement effects along the third direction. It has been shown that the statistics of spatial extremes of the vorticity can depend on the form of the large scale friction that is used in the two-dimensional Navier-Stokes equation, with quadratic drag leading to a different distribution of the spatial extremes as compared to the linear drag \cite{tsang2010nonuniversal}. We concentrate only on the case of linear friction and study the effect of large scale friction and viscosity on the statistics of extreme velocity gradients. Two-dimensional turbulence leads to the formation of a large-scale condensate in the absence of friction. Strong enough friction can dissipate or break down the condensate mode leading to multiple vortex states. The presence or absence of large scale condensates will be shown to alter the statistics of such spatial extremes.

In this work we obtain the distribution of spatial extremes of the vorticity and strain rate tensor as a function of the dissipation parameters using numerical simulations. Section \ref{sec:setup} explains the problem setup, the governing equation and the quantities under study. Section \ref{sec:distributions} discusses the distribution of the spatial extremes, its moments and the dependence on dissipation parameters. In Section \ref{sec:trunc_eul}, using the truncated Euler formalism we study the distribution of the extremes for the non-dissipative equations. In Section \ref{sec:correlation} the power spectrum of the fields is constructed to study the difference in the distributions arising from the dissipative Navier-Stokes and the truncated Euler equations. 

\section{\label{sec:setup} Problem setup}

We consider the two-dimensional incompressible Navier-Stokes equation in a double periodic domain of dimension $[0, L] \times [0, L]$. The governing equation written in terms of the stream function $\psi$ is written as
\begin{align}
\partial_t {\bm \nabla}^2 \psi + \left( {\bm \nabla} \times (\psi \hat{\bf e}_z ) \right) \cdot {\bm \nabla}  \nabla^2 \psi = \nu {\bm \nabla}^4 \psi - \mu {\bm \nabla}^2 \psi +{\bm \nabla}^2  f_{\psi}.
\end{align}
Here the velocity field ${\bf u}$ is given by ${\bf u} = {\bm \nabla} \times (\psi \hat{\bf e}_z ) = u_1 {\bf e}_x + u_2 {\bf e}_y$, $f_\psi$ is the external forcing and $\mu$ is the bottom friction coefficient. The vorticity is then given by $\omega = - {\nabla}^2 \psi$. We define the Reynolds number $\Rey$, the large scale Reynolds number $\Rh$ by,
\begin{align}
\Rey = \frac{U L}{\nu}, \quad 
\Rh = \frac{U}{\mu L}.
\end{align}
Here $U$ is the rms value of the velocity field defined as, $U =  \sqrt{\left\langle |{\bf u}|^2 \right\rangle}$ where the symbol $\left\langle \cdot \right\rangle$ denotes the spatial and temporal averaging. The forcing is chosen to be the standard Kolmogorov forcing defined as,
\begin{align}
f_{\psi} = f_0 \sin (k_f y).
\end{align}
The non-dimensional forcing wave-number is kept constant for the entire study at $k_f L = 8 \pi$. We define the injected energy as $\epsilon = \left\langle {\bf f} \cdot {\bf u} \right\rangle$ which at statistical steady state is given by $\epsilon = \nu \left\langle \omega^2 \right\rangle + \mu U^2$. The equations were integrated using a Fourier pseudo-spectral method and the time marching was done using the ARS-443 scheme \cite{ascher1997implicit} with a CFL condition controlling the time-step at every iteration. 

The spatial extremes of the vorticity $\omega^{\text{min}}, \omega^{\text{max}}$ over the spatial co-ordinates are defined as,
\begin{align}
\omega^{\text{min}} (t) = \min_{x, y} \omega (x, y, t), \quad \quad \omega^{\text{max}} (t) = \max_{x,y} \omega (x, y, t) \label{eqn:extreme_def}
\end{align}
Due to symmetry of the forcing, and the domain, the distributions of the spatial maxima is the same as the distribution of the spatial minima of the vorticity field. We also define the components of the rate of strain rate tensor $S_{ij} = (\partial_i u_j + \partial_j u_i)/2$. The invariants under rotations of the strain rate tensor in two-dimensions are the trace  and the determinant. The trace of the strain rate tensor is zero by incompressibility, while the determinant of the strain rate tensor is given by $S_d (x, y, t) = - ( (\partial_x u_1)^2 + (\partial_x u_2 + \partial_y u_1)^2/4)$. The local viscous dissipation rate at a point is related to the determinant of the strain rate tensor and is given by $-4 \nu S_d$. We study the minimum of the determinant of the strain rate tensor denoted by, 
\begin{align}
S_d^{\text{min}} (t) & = \min_{x,y} S_d (x, y, t)
\end{align}
which also corresponds to the spatial maxima of the local viscous dissipation rate. 

We conduct numerical simulations to study the distribution of extremes as a function of the control parameters $\Rey, \Rh$. Additional studies are carried out to ensure that the distributions of the extremes are independent of the discretisation and time stepping, see Appendix \ref{App:app_1}. The study of spatial extremes is computationally expensive, to obtain converged distributions we have to integrate the equations for very long time. We integrate the system upto and beyond the viscous time scales $L^2/\nu$, which can also be written as the Reynolds number $\Rey$ times the turn over time scale $L/U$. To verify that the distribution $P(X^{\text{min}})$ of the variable $X^{\text{min}}$ is converged we compute the first four moments of the variable $X^{\text{min}}$. We consider the distribution converged if the first four moments computed from half the time series do not vary by more than $5\%$ as compared to the ones computed with the full time series.  



Table \ref{tab:table1} shows the set of parameters for which the distribution of extremes were studied. The first set of simulations (Runs 1-5) were carried out at increasing values of $\Rey$ in the absence of large scale friction $\Rh = \infty$ or $\mu = 0$. While the second set of runs were carried out by setting $\Rey$ approximately constant and increasing value of $\Rh$ to study the influence of the large scale friction on the distribution of the spatial extremes of vorticity and determinant of the strain rate tensor (Runs 6-9). The final set of runs (Runs 10-12) are carried out to verify whether the distributions obtained are independent of the resolution and CFL conditions. The durations of the simulations in the statistical steady state is recorded in the table as $T U/L$ which is the number of turn-over times the simulation has run. As seen from the values in Table \ref{tab:table1} the simulations are integrated over long times, upto viscous time scales. The final column of the table \ref{tab:table1} shows the ratio between the viscous dissipation scale $\eta$ over the smallest length scale of the simulation given by the cut-off wavenumber $k_c$. $\eta$ is estimated by assuming a forward cascade of enstrophy which leads to $\eta = \ell_f \Rey^{-1/2}$ where $\ell_f = 2 \pi/k_f$. $k_c \eta$ having value larger than $1$ then implies that the viscous scales are well resolved.


\begin{table*}
\caption{\label{tab:table1} Table shows the parameters for different two-dimensional Navier-Stokes simulations that were carried out in this study. The duration of the simulation is denoted by $T \frac{U}{L}$, $k_c \eta$ gives an estimate of the ratio between the dissipation scale and the smallest length scale that is resolved. }
\begin{ruledtabular}
\begin{tabular}{cccccccc}
 & Resolution &$\Rey $  & $\Rh$ &$T \frac{U}{L}$ & CFL & $k_c \eta$ \\
\hline
Run 1   & $1024 \times 1024$ & $3.2 \times 10^4$ & $-$ & $6.7 \times 10^4$ & $0.5$ & $3.0$  \\ 
Run 2   & $512 \times 512$ & $1.2 \times 10^4$ & $- $& $8.1 \times 10^4$ & $0.5$ & $2.5$  \\ 
Run 3   & $384 \times 384$ & $3.2 \times 10^3$ & $-$ & $7.6 \times 10^4$ & $0.5$ & $3.5$  \\ 
Run 4   & $256 \times 256$ & $1.2 \times 10^3$ & $-$ & $6.1 \times 10^4$ & $0.5$ & $3.8$  \\ 
Run 5   & $256 \times 256$ & $4.2 \times 10^2$ & $-$ & $4.7 \times 10^4$ & $0.5$ & $6.5$ \\
\hline
Run 6   & $1024 \times 1024$ & $3.2 \times 10^4$ & $4.0 \times 10^3$ & $4.7 \times 10^4$ & $0.5$ & $2.9$  \\ 
Run 7   & $1024 \times 1024$  & $3.3 \times 10^4$ & $2.9 \times 10^2$ & $5.1 \times 10^4$ & $0.5$ & $3.0$ \\ 
Run 8   & $1024 \times 1024$ & $3.5 \times 10^4$ & $1.7 \times 10^1$ & $3.7 \times 10^4$ & $0.5$ & $2.9$  \\ 
Run 9 & $1024 \times 1024$ & $3.5 \times 10^4$ & $9.7 \times 10^{-1}$ & $2.5 \times 10^4$ & $0.5$ & $2.9$ \\ 
\hline
Run 10 & $768 \times 768$ & $3.2 \times 10^4$ & $-$ & $9.5 \times 10^3$ & $0.25$ & $2.2$ \\
Run 11 & $1536 \times 1536$ & $3.2 \times 10^4$ & $-$ & $2.3 \times 10^3$ & $0.25$ & $4.5$ \\
Run 12 & $768 \times 768$ & $3.2 \times 10^4$ & $-$ & $9.4 \times 10^3$ & $0.1$ & $2.2$ \\
\end{tabular}
\end{ruledtabular}
\end{table*}


\section{\label{sec:distributions} Distributions of the extrema}

\begin{figure*}
     \centering
     \begin{subfigure}[b]{0.325\textwidth}
         \centering
         \includegraphics[width=\textwidth]{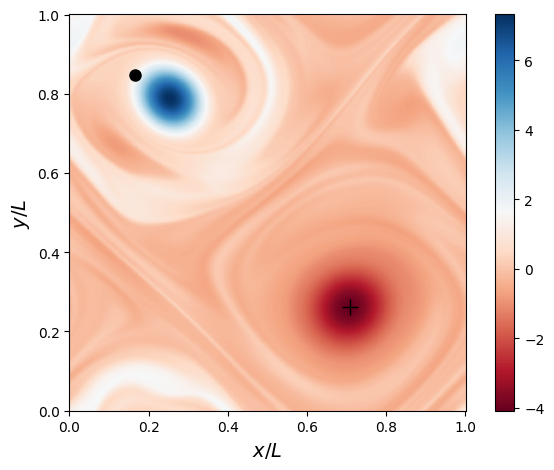}
         \caption{}
         \label{fig:flow_Re_1}
     \end{subfigure}
     \hfill
     \begin{subfigure}[b]{0.325\textwidth}
         \centering
         \includegraphics[width=\textwidth]{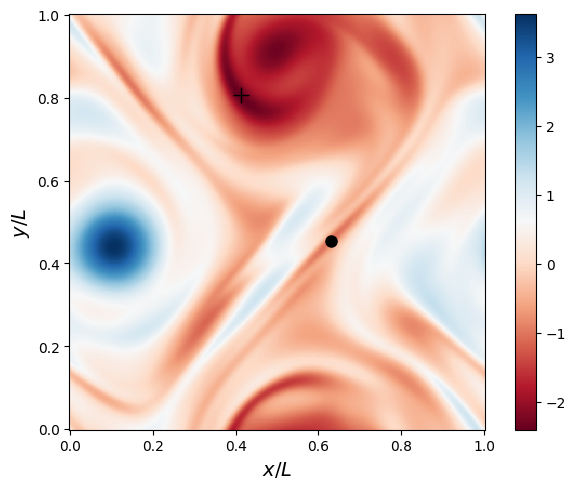}
         \caption{}
         \label{fig:flow_Re_2}
     \end{subfigure}
     \hfill
     \begin{subfigure}[b]{0.325\textwidth}
         \centering
         \includegraphics[width=\textwidth]{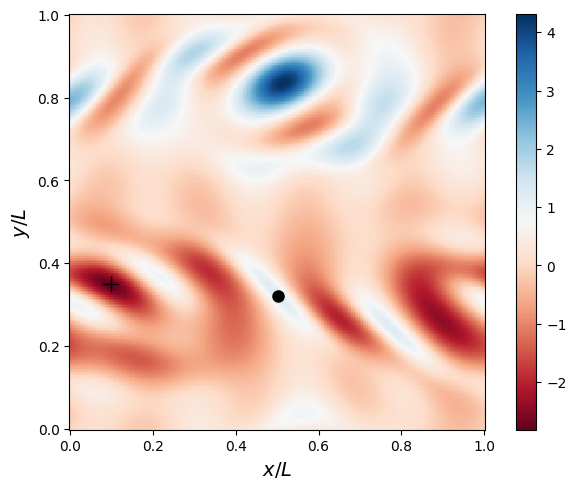}
         \caption{}
         \label{fig:flow_Re_3}
     \end{subfigure} \\
     \begin{subfigure}[b]{0.325\textwidth}
         \centering
         \includegraphics[width=\textwidth]{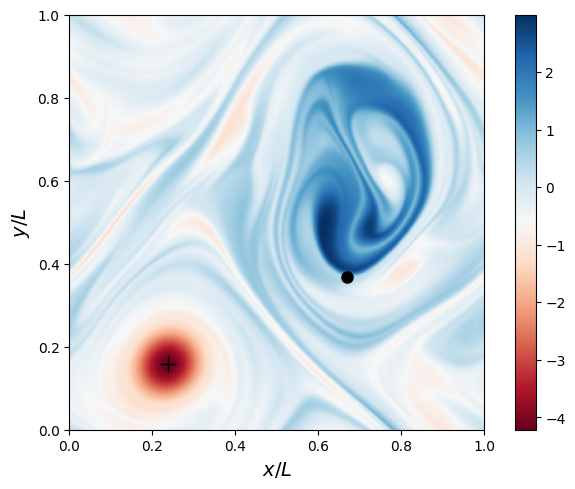}
         \caption{}
         \label{fig:flow_Rh_1}
     \end{subfigure}
\begin{subfigure}[b]{0.325\textwidth}
         \centering
         \includegraphics[width=\textwidth]{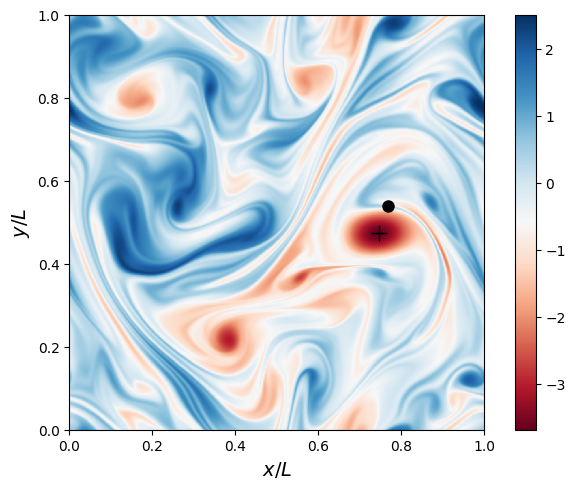}
         \caption{}
         \label{fig:flow_Rh_2}
     \end{subfigure}
     \begin{subfigure}[b]{0.325\textwidth}
         \centering
         \includegraphics[width=\textwidth]{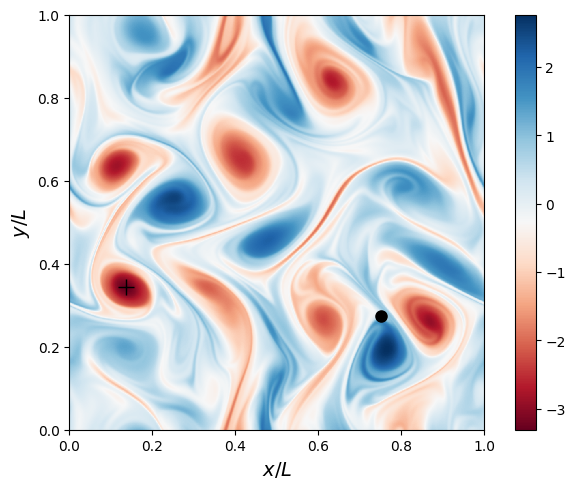}
         \caption{}
         \label{fig:flow_Rh_3}
     \end{subfigure}
        \caption{Figures show the vorticity field plots along with the spatial location of the minima of the fields $\omega$ and $S_{d}$.  The top row shows the contour plots for Runs a) 1, b) 3 and c) 5, while the bottom row shows for Runs d) 7, e) 8 and f) 9, see Table \ref{tab:table1} for the parameters corresponding to these runs.  The markers $\Plus$, 
        and \mytikzdot    
         denote the locations of $\omega^{\text{min}}, S_{d}^{\text{min}}$ respectively. }
        \label{fig:flow_Re_Rh}
\end{figure*}

Figures \ref{fig:flow_Re_Rh} show the visualisation of the vorticity of the two-dimensional turbulent flow at an instant of time along with the spatial location of the extremes. The figures correspond to different values of $\Rey$ and $\Rh$ for the parameters a) $\Rey = 3.2 \times 10^4$, b) $\Rey = 3.2 \times 10^3$, c) $\Rey = 4.2 \times 10^2$, d) $\Rey = 3.3 \times 10^4, \Rh = 2.9 \times 10^2$, e) $\Rey = 3.5 \times 10^4, \Rh = 1.7 \times 10^1$ and f) $\Rey = 3.5 \times 10^4, \Rh = 0.97$. The marker $\Plus$ denotes the spatial location of $\omega^{\text{min}}$, while the marker \tikz\draw[black,fill=black] (0,0) circle (.5ex); denotes the spatial location of $S_{d}^{\text{min}}$. For the cases where both $\Rey, \Rh$ are large, the minimum of vorticity $\omega^{\text{min}}$ is found to be at the core of the large scale negative vortex for all time instances that we have observed. The minimum of the determinant of the strain rate tensor is found to be present away from the center of the vortices, either near the edges of the vortices or in the shear layers in between vortex regions. As $\Rh, \Rey$ values are decreased the large scale condensates loose strength and the flow starts to have multiple vortices of comparable strengths. The minimum values of $\omega^{\text{min}}$ are found to be at the center of one of the many vortices and $S_{d}^{\text{min}}$ are found to be in one of the many regions between the vortices.

\begin{figure*}
     \centering
     \begin{subfigure}[b]{0.45\textwidth}
         \centering
         \includegraphics[width=\textwidth]{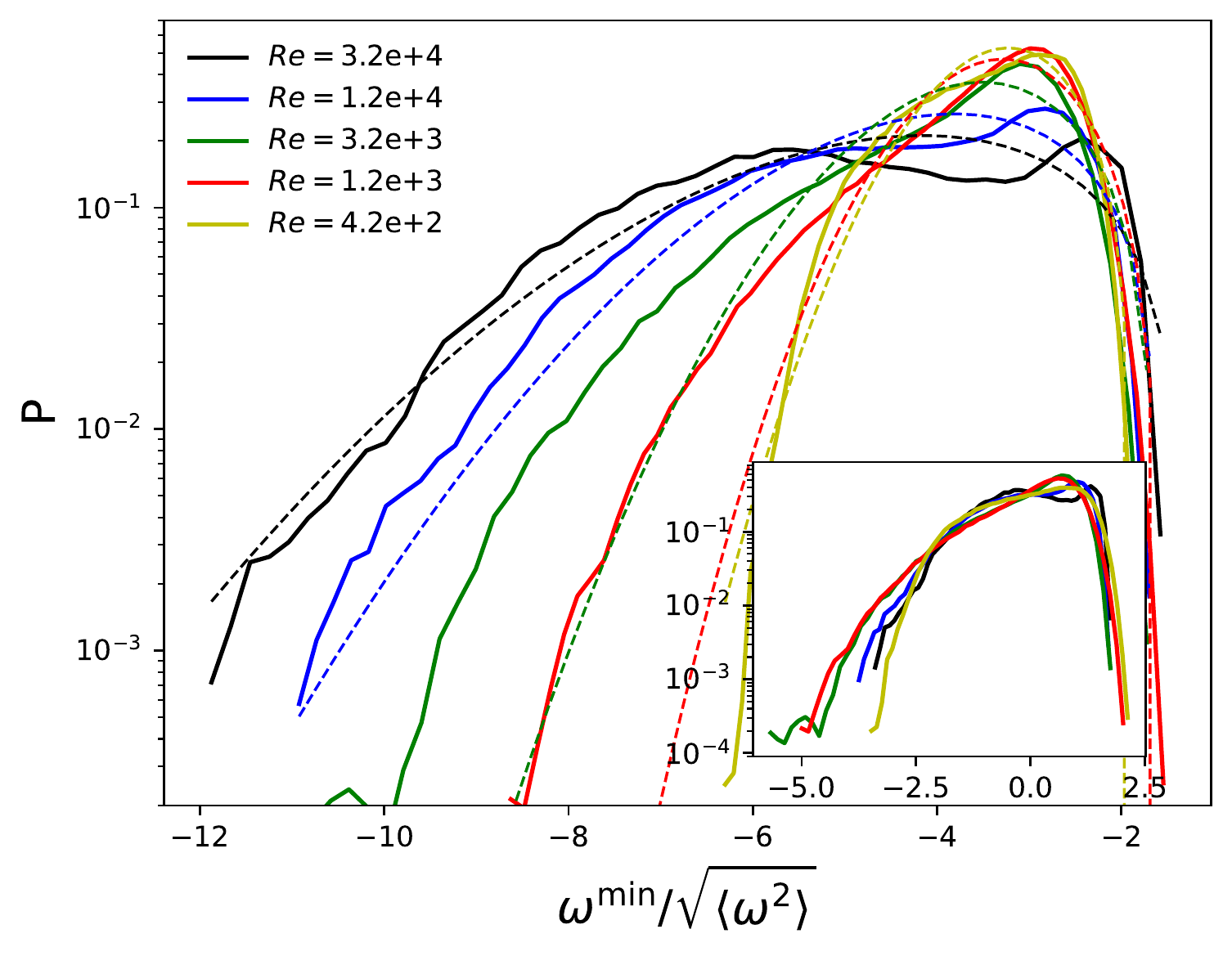}
         \caption{}
         \label{fig:dist_omega_min_Re}
     \end{subfigure}
     \hfill
     \begin{subfigure}[b]{0.45\textwidth}
         \centering
         \includegraphics[width=\textwidth]{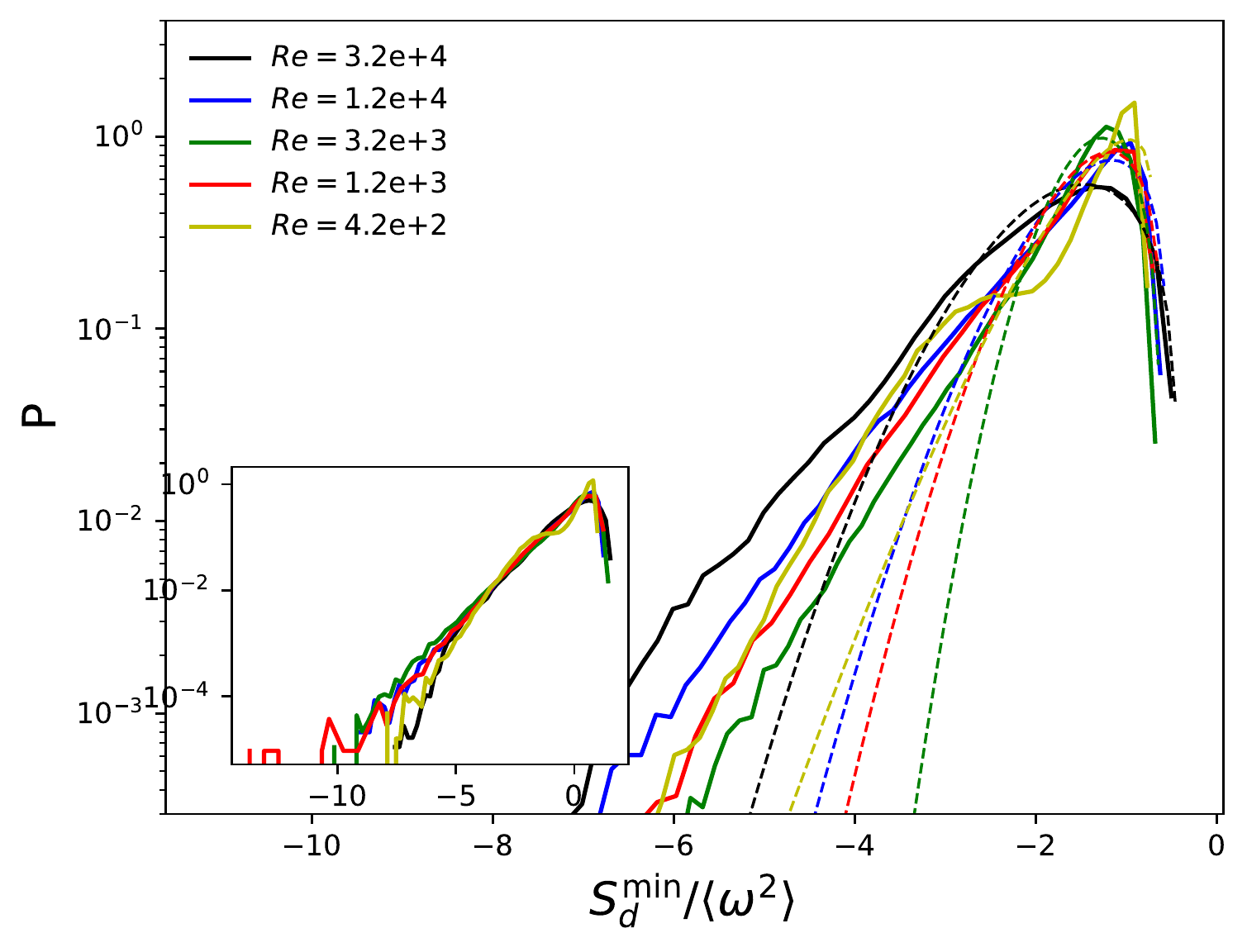}
         \caption{}
         \label{fig:dist_Sd_min_Re}
     \end{subfigure}
        \caption{Figures show the distributions of the normalized a) $\omega^{\text{min}}$, b) $S_{d}^{\text{min}}$ for different values of the $\Rey$. The normalization is done with the rms value of the vorticity $\sqrt{\left\langle \omega^2 \right\rangle}$ and the enstrophy $\left\langle \omega^2 \right\rangle$. The dashed line shows extreme value fits (see equation \eqref{eqn:GEV_dist} ) for the different distributions. Insets show the same distributions without the mean and rescaled with their respective standard deviations. }
        \label{fig:three_graphs_Re}
\end{figure*}

\begin{figure*}
     \centering
     \begin{subfigure}[b]{0.45\textwidth}
         \centering
         \includegraphics[width=\textwidth]{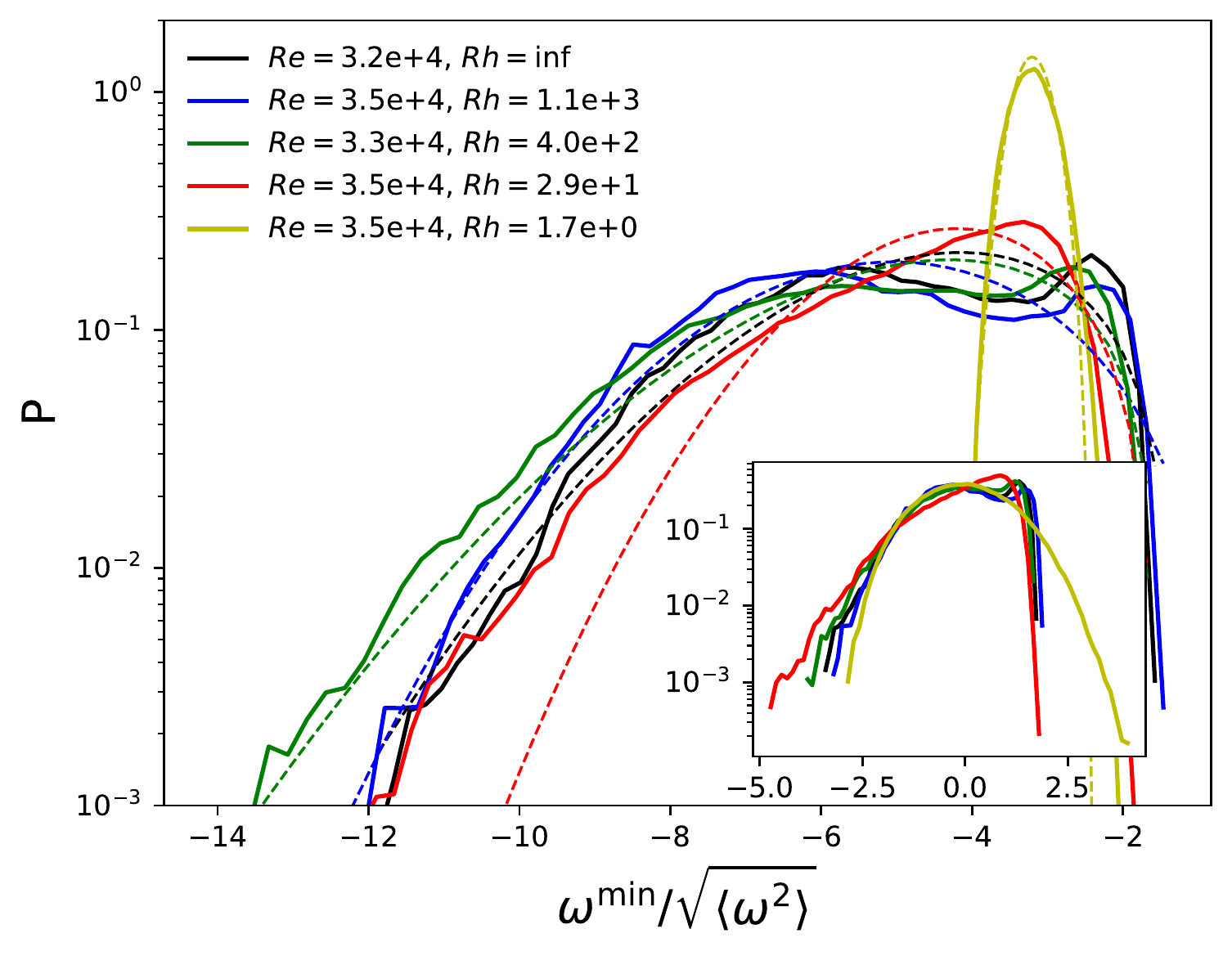}
         \caption{}
         \label{fig:dist_omega_min_Rh}
     \end{subfigure}
     \hfill
     \begin{subfigure}[b]{0.45\textwidth}
         \centering
         \includegraphics[width=\textwidth]{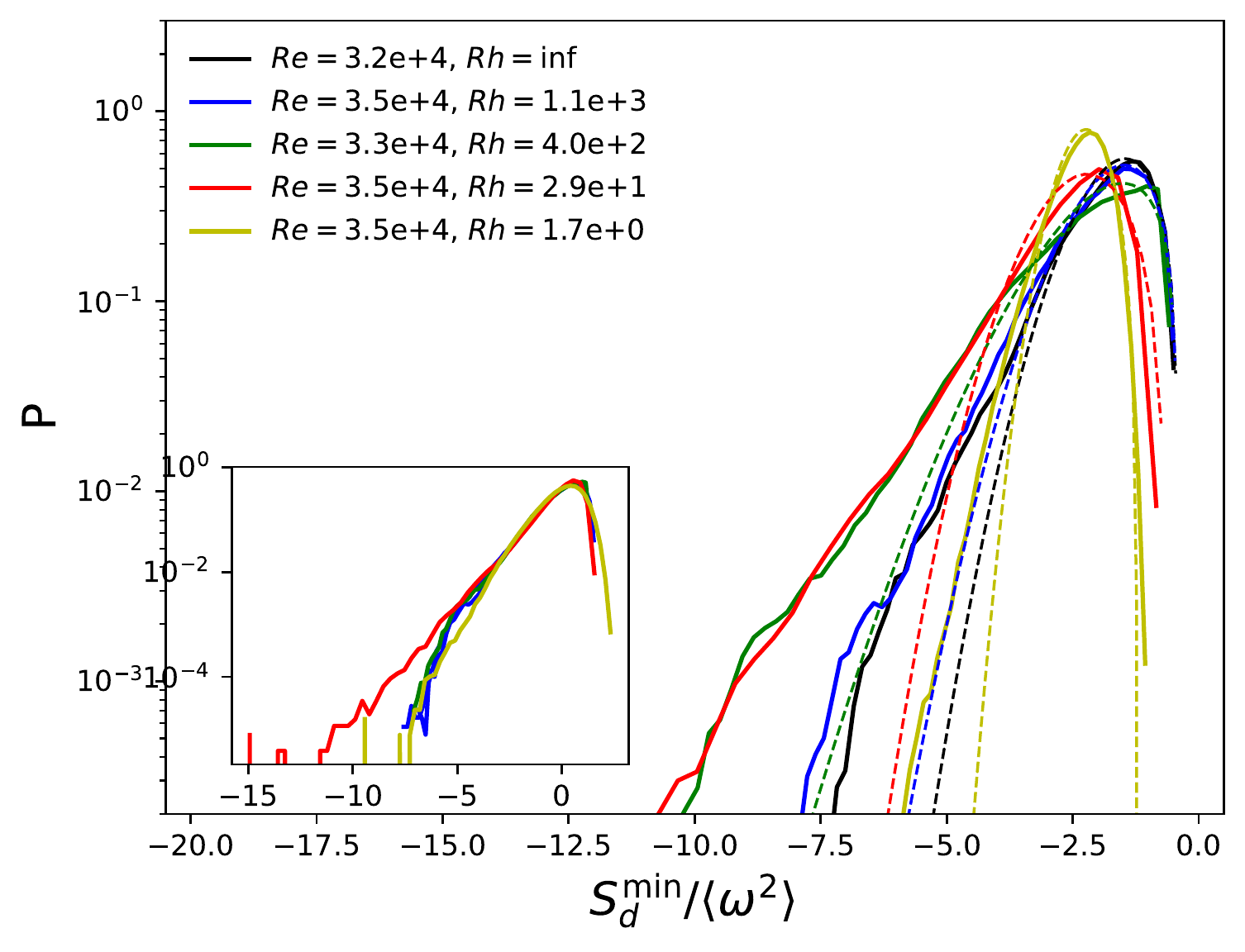}
         \caption{}
         \label{fig:dist_Sd_min_Rh}
     \end{subfigure}
        \caption{Figures show the distributions of the normalized a) $\omega^{\text{min}}$, b) $S_{d}^{\text{min}}$ for different values of the $\Rh$ and at a constant $\Rey$. The normalization is done with the rms value of the vorticity $\sqrt{\left\langle \omega^2 \right\rangle}$ and the enstrophy $\left\langle \omega^2 \right\rangle$. The dashed lines show Generalised extreme value fits (see equation \eqref{eqn:GEV_dist}) for the different distributions.}
        \label{fig:three_graphs_Rh}
\end{figure*}

The PDFs of $\omega^{\text{min}}, S_{d}^{\text{min}}$ are shown in Figs. \ref{fig:three_graphs_Re} for different $\Rey$ numbers. The quantities are normalised with the RMS value of the vorticity $\sqrt{\left\langle \omega^2 \right\rangle}$ and enstrophy $\left\langle \omega^2 \right\rangle$. The inset of the figures show the distributions of the variables $\omega^{\text{min}}, S_{d}^{\text{min}}$ without the mean and rescaled by their respective standard deviations. The distributions are fitted with a Generalised Extreme value (GEV) distribution for minima given by,
\begin{align}
P(x; \xi, \mu, \sigma) & =   \frac{1}{\sigma} \left( 1 + \xi \frac{x- \mu}{\sigma} \right)^{\left( \frac{1}{\xi} - 1 \right)} \times \nonumber \\
& \exp \left( - \left(1 + \xi \frac{x- \mu}{\sigma} \right)^{\frac{1}{\xi}} \right), \label{eqn:GEV_dist}
\end{align}
where $\xi$ is called the shape parameter, $\mu$ the location parameter and $\sigma$ the scale parameter. For $\xi = 0$, the distribution becomes, 
\begin{align}
P(x; \mu, \sigma) = \frac{1}{\sigma} \exp \left( \frac{x- \mu}{\sigma} \right) \exp \left( -\exp \left( \frac{x- \mu}{\sigma} \right)\right).
\end{align}
which corresponds to the Gumbel class of distribution. Depending on the shape parameter $\xi$, the GEV falls into one of Gumbel, Fr{\'e}chet and Weibull distributions. The black dashed line in the figures shows the GEV fit for each of the $\Rey$ curves found using standard python libraries. 
We see that for all the values of $\Rey$, the distributions fall outside the GEV class and from the inset we see that the distributions have not yet reached a $\Rey$ or viscosity independent form even at the highest value we have explored in this study, $\Rey \sim 3.2 \times 10^4$.  
To understand the dependence of these extreme value statistics on the large scale friction, we look at the Runs 6-9 that corresponds to varying $\Rh$ while keeping the $\Rey$ close to constant, see Table \ref{tab:table1}. Figures \ref{fig:three_graphs_Rh} show the PDFs of $\omega^{\text{max}}, S_{d}^{\text{max}}$, non-dimensionalized by $\sqrt{\left\langle \omega^2 \right\rangle}$ and $\left\langle \omega^2 \right\rangle$ respectively. The dashed lines indicate the GEV fits (eq. \eqref{eqn:GEV_dist}) for the PDFs. We see that the presence of a strong large scale friction modifies the distribution of the extremes as compared to the case without any large scale friction. 

\begin{figure*}
     \centering
     \begin{subfigure}[b]{0.45\textwidth}
         \centering
         \includegraphics[width=\textwidth]{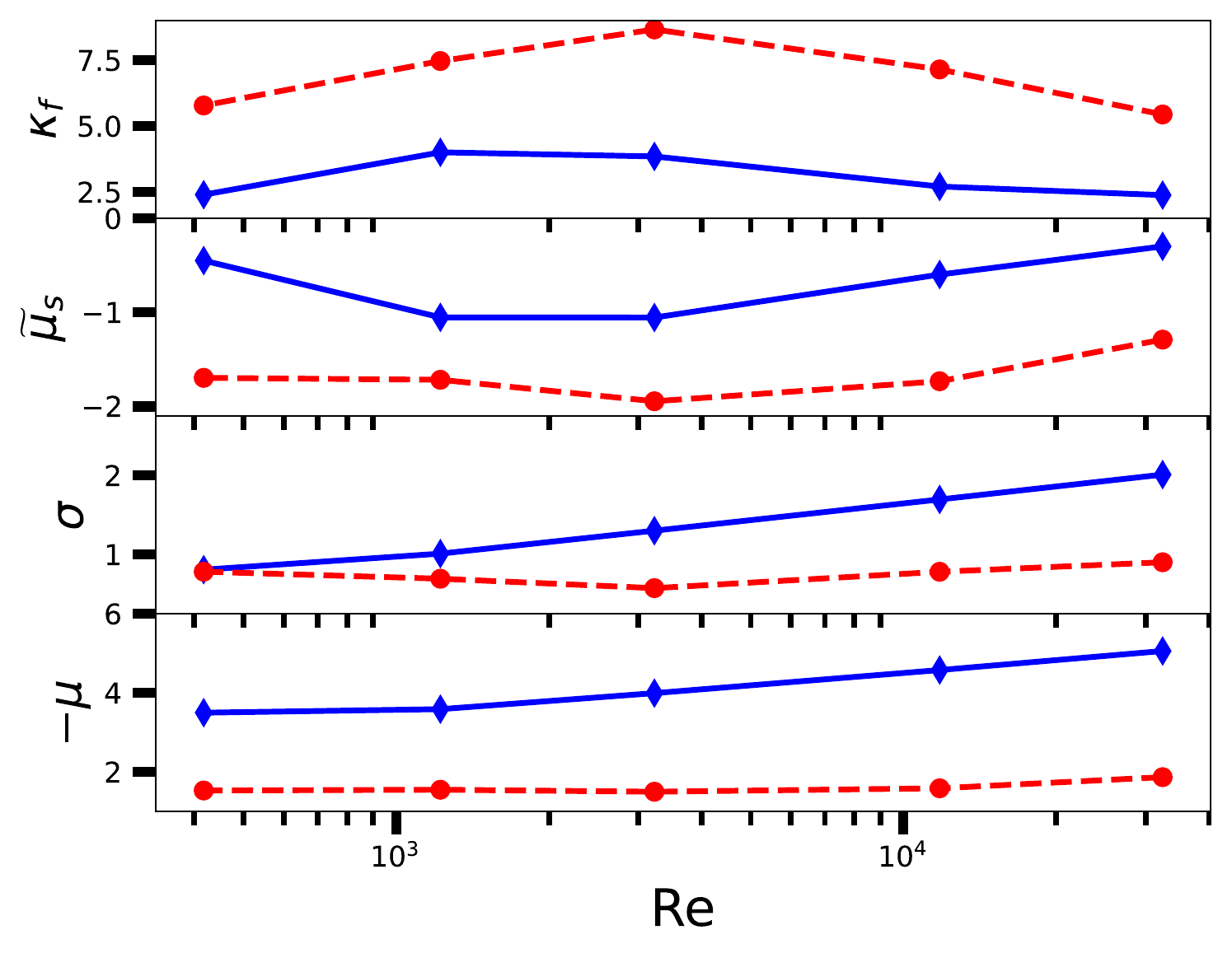}
         \caption{}
         \label{fig:first_Re}
     \end{subfigure}
     \hfill
     \begin{subfigure}[b]{0.45\textwidth}
         \centering
         \includegraphics[width=\textwidth]{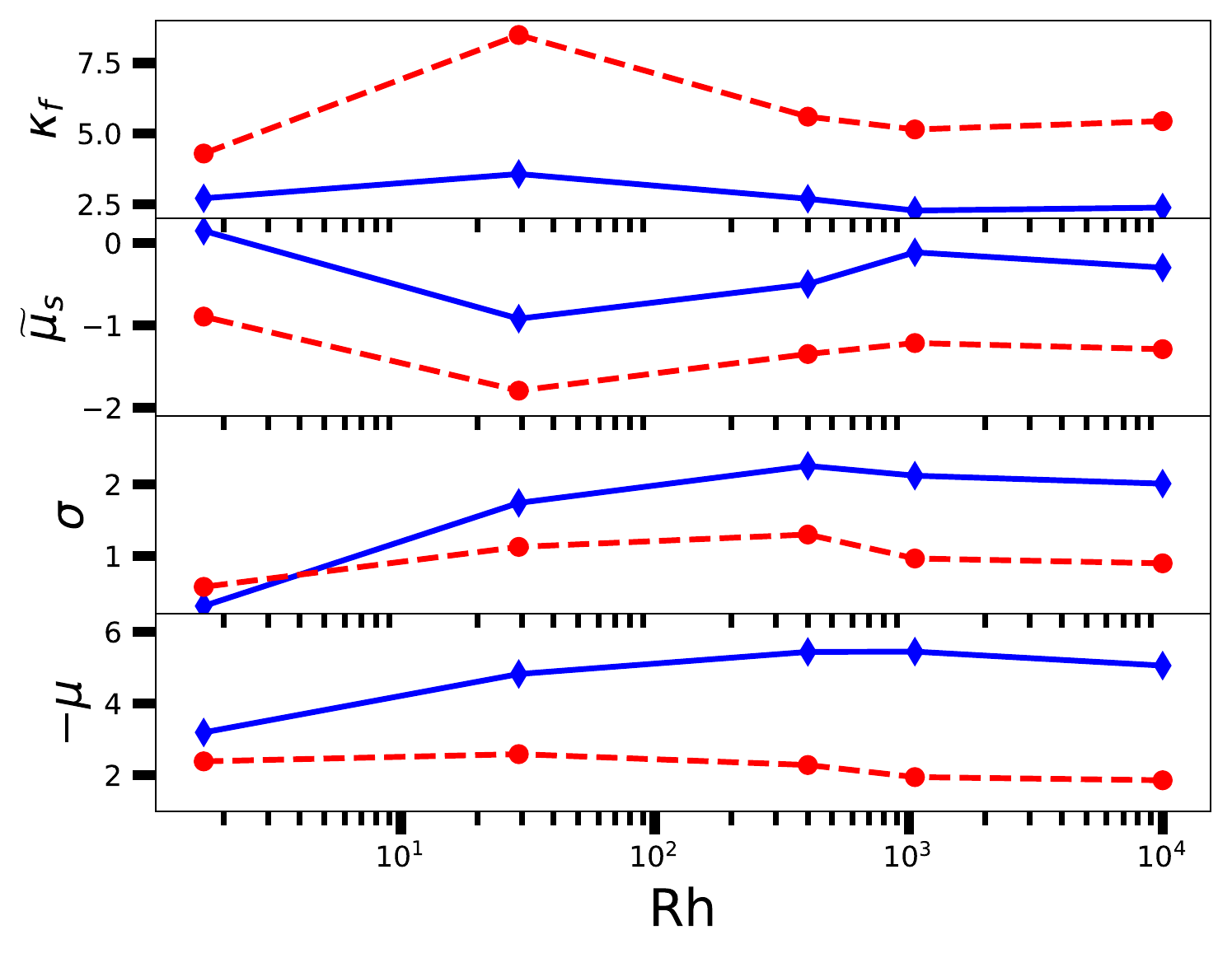}
         \caption{}
         \label{fig:first_Rh}
     \end{subfigure}
        \caption{Figures show the mean $\mu$, standard deviation $\sigma$, skewness $\widetilde{\mu}_s$ and the kurtosis $\kappa_f$ of the distributions of $\omega^{\text{min}}/\sqrt{\left\langle \omega^2 \right\rangle}$ in solid lines and $S_{d}^{\text{min}}/\left\langle \omega^2 \right\rangle$ in dashed lines for different values of a) $\Rey$ and b) $\Rh$. }
        \label{fig:first_std}
\end{figure*}

Figures \ref{fig:first_Re},  \ref{fig:first_Rh} show the mean $\mu$, standard deviation $\sigma$, skewness $\widetilde{\mu}_s$ and the kurtosis $\kappa_f$ of the non-dimensionalised distributions of, $\omega^{\text{min}}$ and $S_{d}^{\text{min}}$ as a function of $\Rey$ and $\Rh$. The mean and standard deviation of these quantities have a weak dependence on $\Rey$, both growing slowly as we increase $\Rey$. Using the method of bounds, it is known that for monochromatically forced two-dimensional turbulence the enstrophy is bounded by a viscosity independent scaling \cite{alexakis2006energy} and the mean of the spatial extremes of vorticity and velocity gradients was shown to be logarithmically growing with $\Rey$ \cite{gallet2015exact}. Presence of strong large scale friction effects ($\Rh$ small) break large vortices into smaller ones, and reduces $\mu$ and $\sigma$ of  $\omega^{\text{min}}, S_{d}^{\text{min}}$.  The skewness of the distributions is always negative indicating that the tails corresponding to more negative values, are fatter. The flatness of the minima of vorticity is smaller than the value obtained for a standard Gaussian distribution while for the determinant of the strain rate tensor we find that the flatness is much larger due to large tails.  

\section{\label{sec:trunc_eul} Truncated Euler}

\begin{figure*}
     \centering
     \begin{subfigure}[b]{0.45\textwidth}
         \centering
         \includegraphics[width=\textwidth]{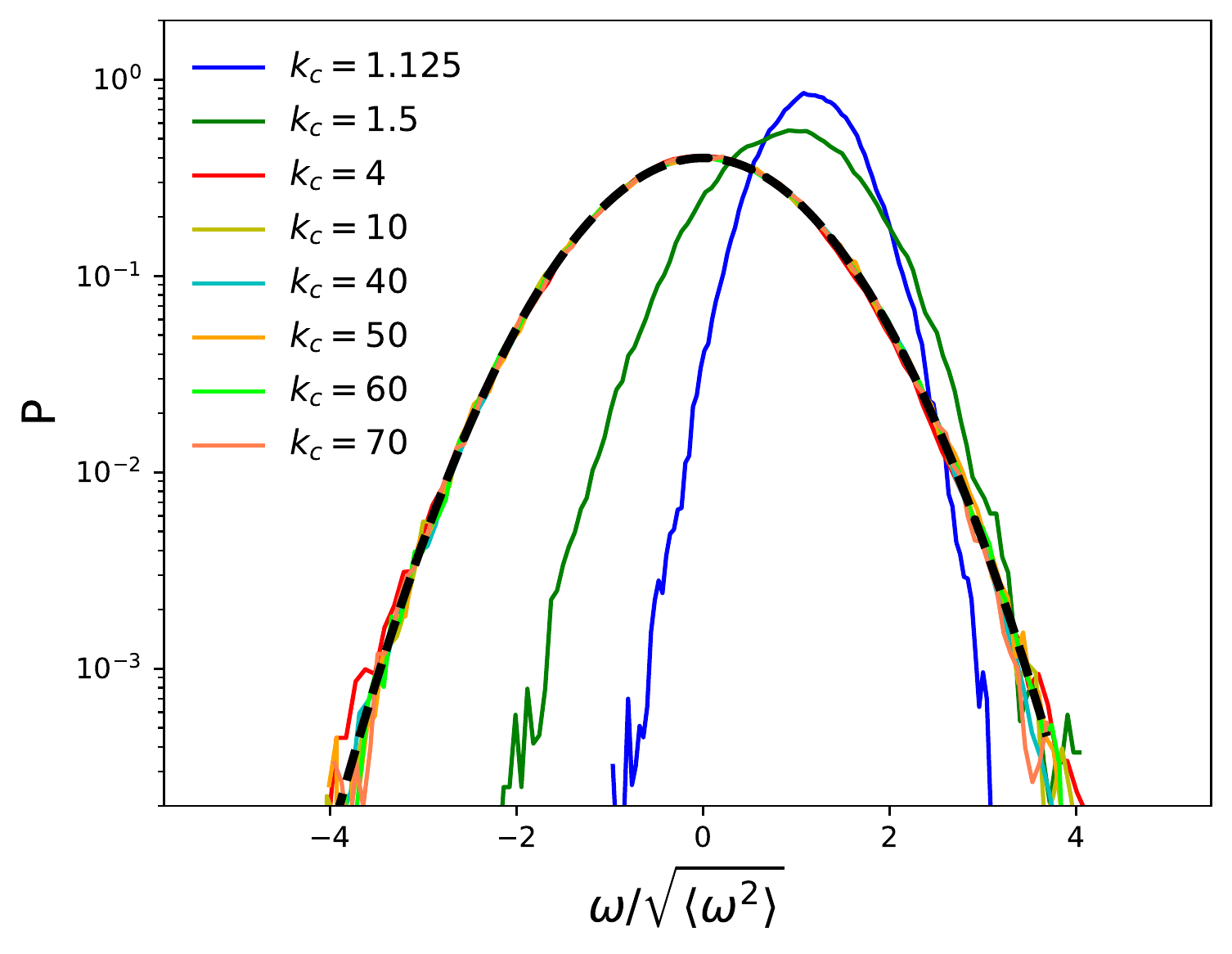}
         \caption{}
         \label{fig:vort_point}
     \end{subfigure}
     \hfill
     \begin{subfigure}[b]{0.45\textwidth}
         \centering
         \includegraphics[width=\textwidth]{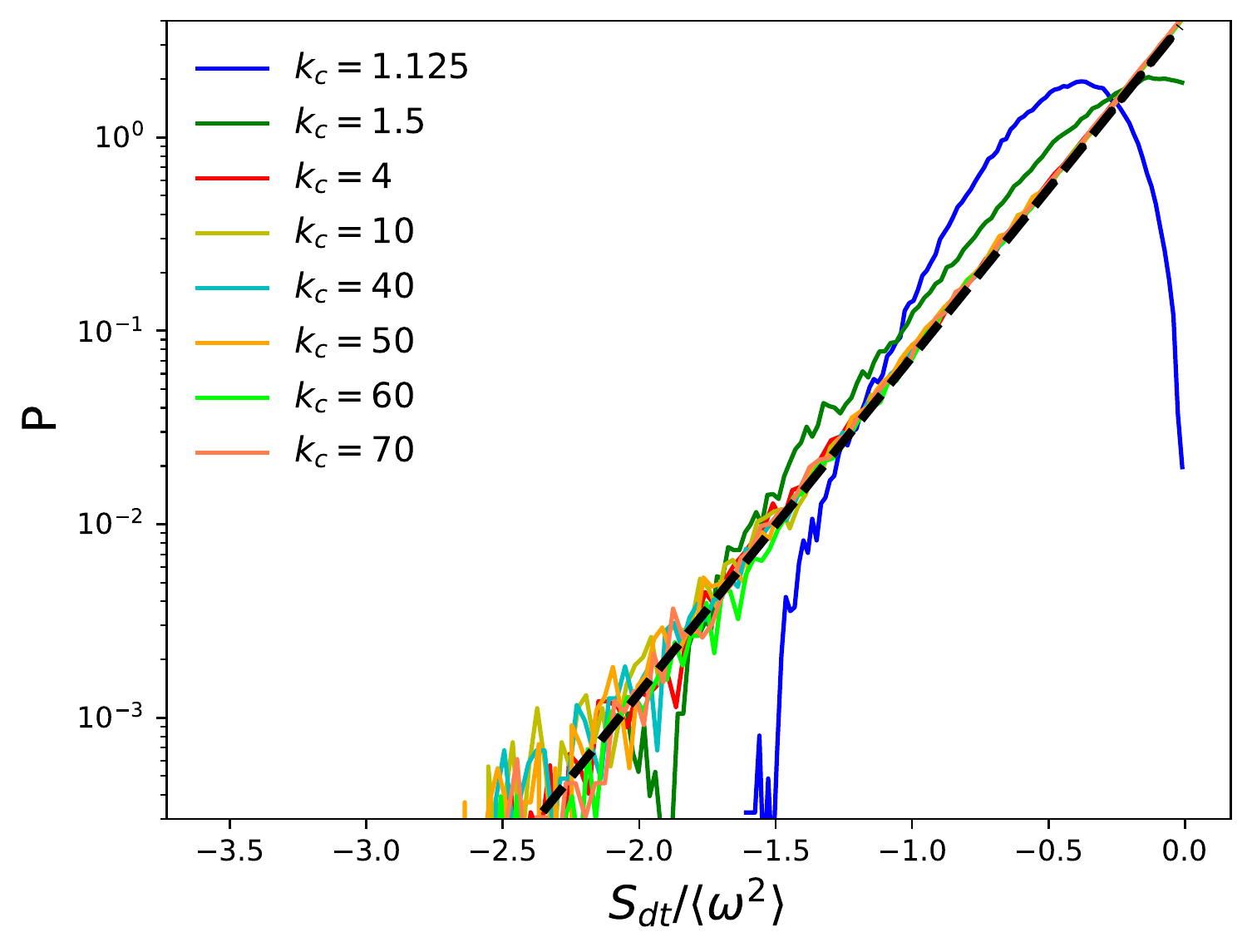}
         \caption{}
         \label{fig:Sdt_point}
     \end{subfigure}
        \caption{Figures show the distributions of $\omega/\sqrt{\left\langle \omega^2 \right\rangle}$ and $S_{d}/\left\langle \omega^2 \right\rangle$ at a fixed spatial location for different parameters $k_c$ as mentioned in the legend. The black dashed line indicates the normalized Gaussian distribution in a) and exponential distribution in b).}
        \label{fig:point_data}
\end{figure*}

A simple model that has been shown in some cases to effectively model the dynamics of $2D$ turbulence is the Truncated Euler Equations (TEE) \cite{kraichnan1975statistical, dallas2020transitions, van2022geometric}. The TEE model is based on the equilibrium solutions of the Navier Stokes equations where the forcing and dissipation terms are absent. The TEE model conserves the two invariants, the energy $E$ and the enstrophy $\Omega$. The dynamics of the TEE model can be understood using tools of equilibrium statistical mechanics, which predicts three different regimes based on the parameter $k_c = \sqrt{\frac{\Omega}{E}}$, \cite{kraichnan1975statistical}. We simulate the truncated Euler equations on a $N_e \times N_e$ grid for different values of $k_c$. We fix the value of $N_e$ to be $N_e = 256$ for all the TEE simulations and the simulations are run for more than a thousand turn over times for each value of $k_c$ explored. In the canonical ensemble description of the truncated Euler equations, we expect that the vorticity at a given spatial location follows a Gaussian distribution $P(\omega) \propto \exp ( - \gamma \omega^2 )$ with $\gamma$ being $1/(2\Omega)$ and the determinant of the strain tensor follows an exponential distribution $P(S_d) \propto \exp(8\gamma S_d)$.  Figure \ref{fig:point_data} shows the distribution of the vorticity and the determinant of the strain rate tensor at a given location $(x, y)$ for different values of $k_c$. For large values of $k_c$ we see that the distributions for the vorticity to be Gaussian and for the determinant of the strain rate tensor to be exponential. For low values of $k_c$, the TEE system is in the condensate phase where the distributions deviate from the Gaussian distribution and a microcanonical description is necessary \cite{van2022geometric}. 

\begin{figure}
     \begin{subfigure}[b]{0.45\textwidth}
         \centering
         \includegraphics[width=\textwidth]{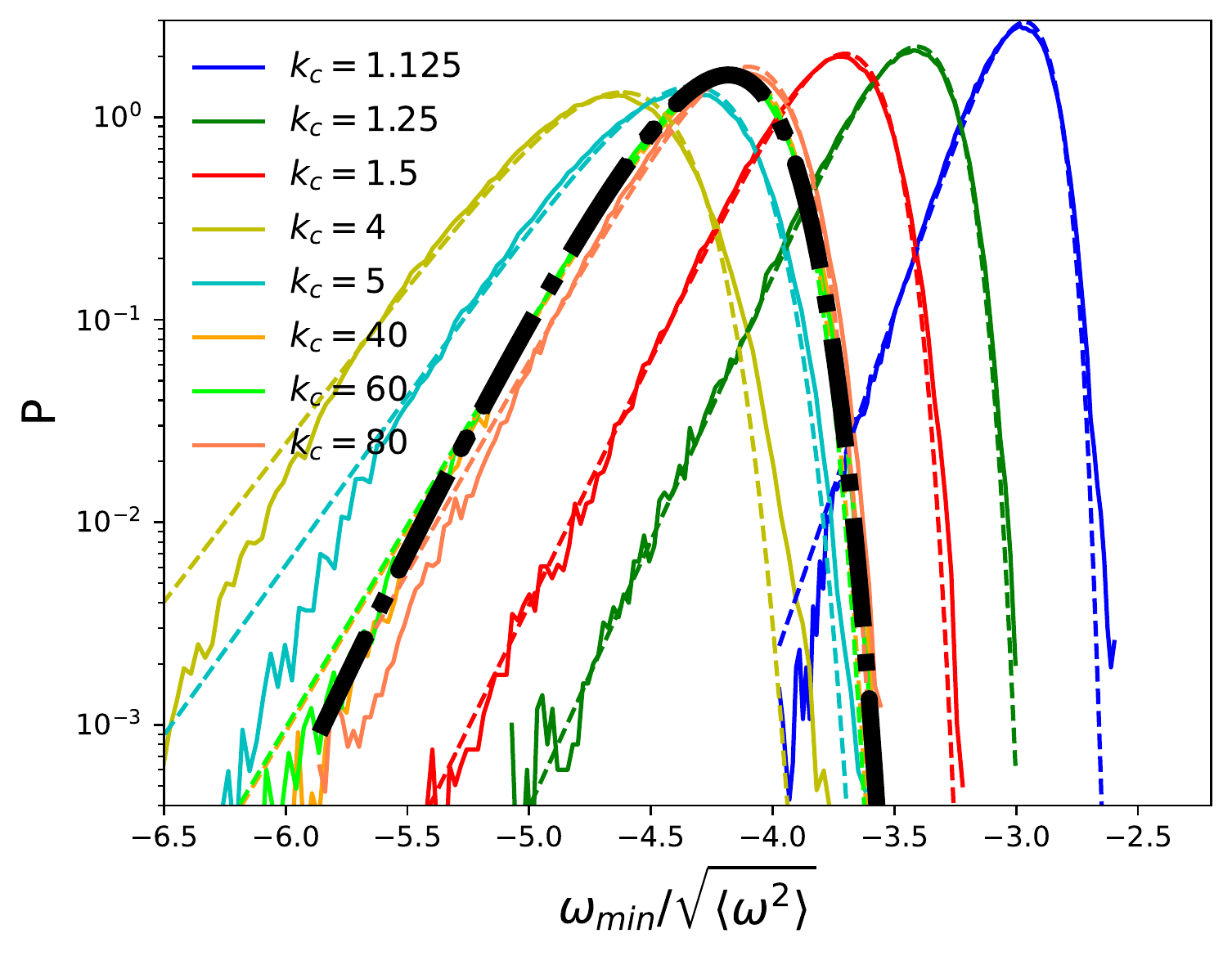}
         \caption{}
         \label{fig:TEE_omega}
     \end{subfigure}
     \begin{subfigure}[b]{0.45\textwidth}
         \centering
         \includegraphics[width=\textwidth]{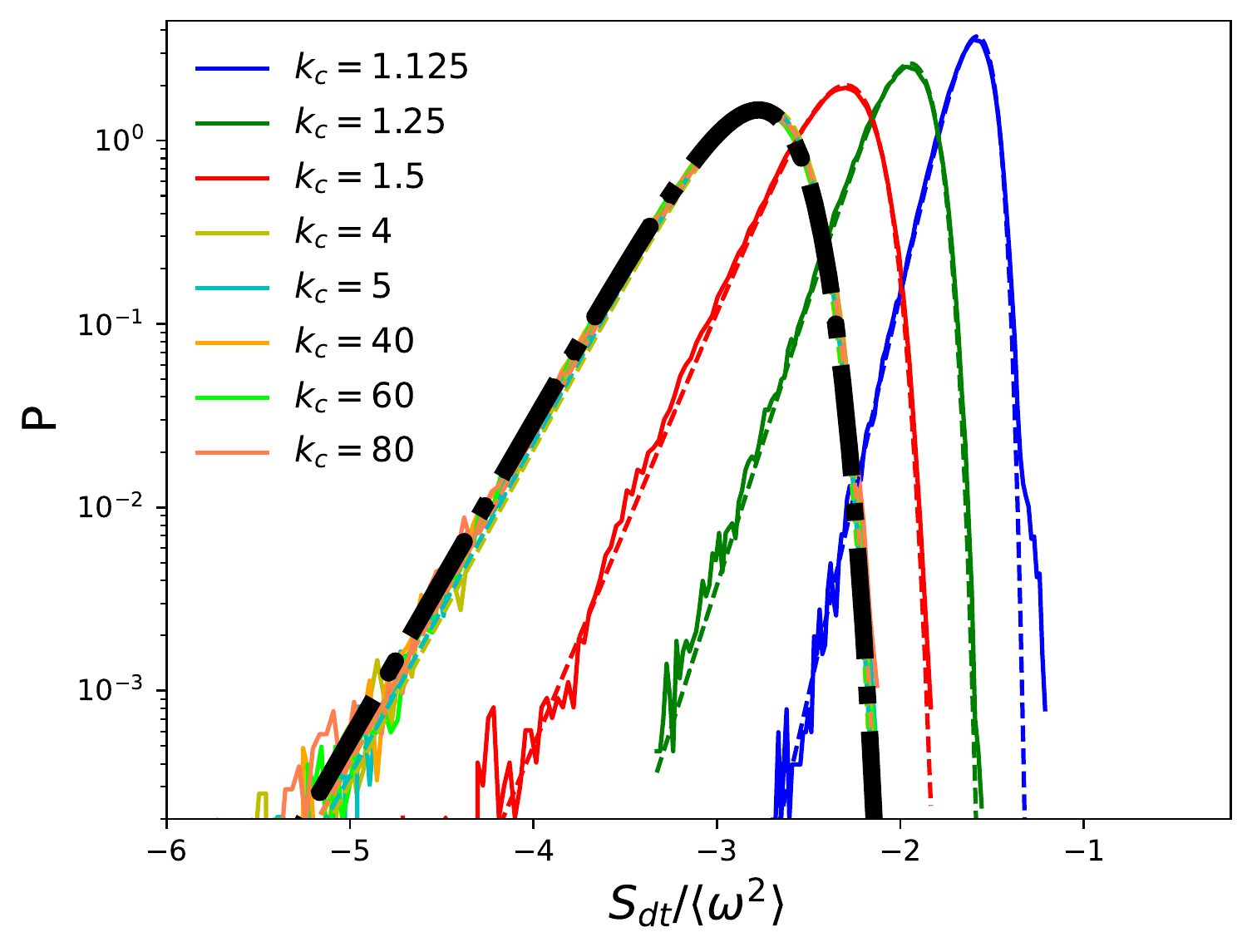}
         \caption{}
         \label{fig:TEE_Sdt}
     \end{subfigure}
\caption{\label{fig:TEE_dist} The figure shows the probability distribution of a) $\omega^{\text{min}}$ and b) $S_{d}^{\text{min}}$ for different values of $k_c$ for the truncated Euler Equations. The dashed lines show the fit obtained using the Gumbel distribution and the black dashed dotted lines show the solutions obtained in eqs. \ref{eqn:sol_wmin},\ref{eqn:sol_Smin}. }
\end{figure}

Assuming that for sufficiently large $k_c$, the vorticity at each point is independent of each other and they follow the distribution in Fig. \ref{fig:point_data}. We can then find the spatial minimum of the vorticity to be given by, 
\begin{eqnarray}
P(\omega^{\text{min}}) = & \frac{N_t}{2^{N_t-1}} \sqrt{ \frac{\gamma}{\pi} } \exp \left( -\gamma( \omega^{\text{min}} )^2 \right) \left(  \right. \nonumber \\
& \left. 1 +  \erf  \left( \sqrt{\gamma} \omega^{\text{min}} \right) \right)^{N_t-1} \label{eqn:sol_wmin}
\end{eqnarray}
where $N_t = N_e^2$ is the total number of grid points in real space used to solve the truncated Euler equations and erf is the error function. Similarly for the determinant of the strain rate tensor we can find the spatial minimum to follow the distribution given by,
\begin{equation}
P(S_d^{\text{min}}) = 8 \gamma N_t \exp (8 \gamma S_d^{\text{min}}) \left( 1 - \exp (8 \gamma S_d^{\text{min}}) \right)^{N_t-1} \label{eqn:sol_Smin}
\end{equation}
Both of the distributions in equations \eqref{eqn:sol_wmin}, \eqref{eqn:sol_Smin}, can be shown to fall under the Gumbel form of the extreme value distribution \cite{majumdar2020extreme}.  We show the distribution of the minima of the vorticity and the determinant of the strain rate tensor in Figs \ref{fig:TEE_dist} for the TEE model for different values of $k_c$. The dashed lines denote the Gumbel fit to these distributions and the thick dashed dotted line denotes the analytical distributions from eqns.  \eqref{eqn:sol_wmin}, \eqref{eqn:sol_Smin}. For values of $k_c$ which are large, where the distributions of the fields at a given spatial location are Gaussian, we get the standard Gumbel distribution for the spatial extremes. 

\section{\label{sec:correlation} Power spectrum}

\begin{figure}
     \begin{subfigure}[b]{0.45\textwidth}
         \centering
         \includegraphics[width=\textwidth]{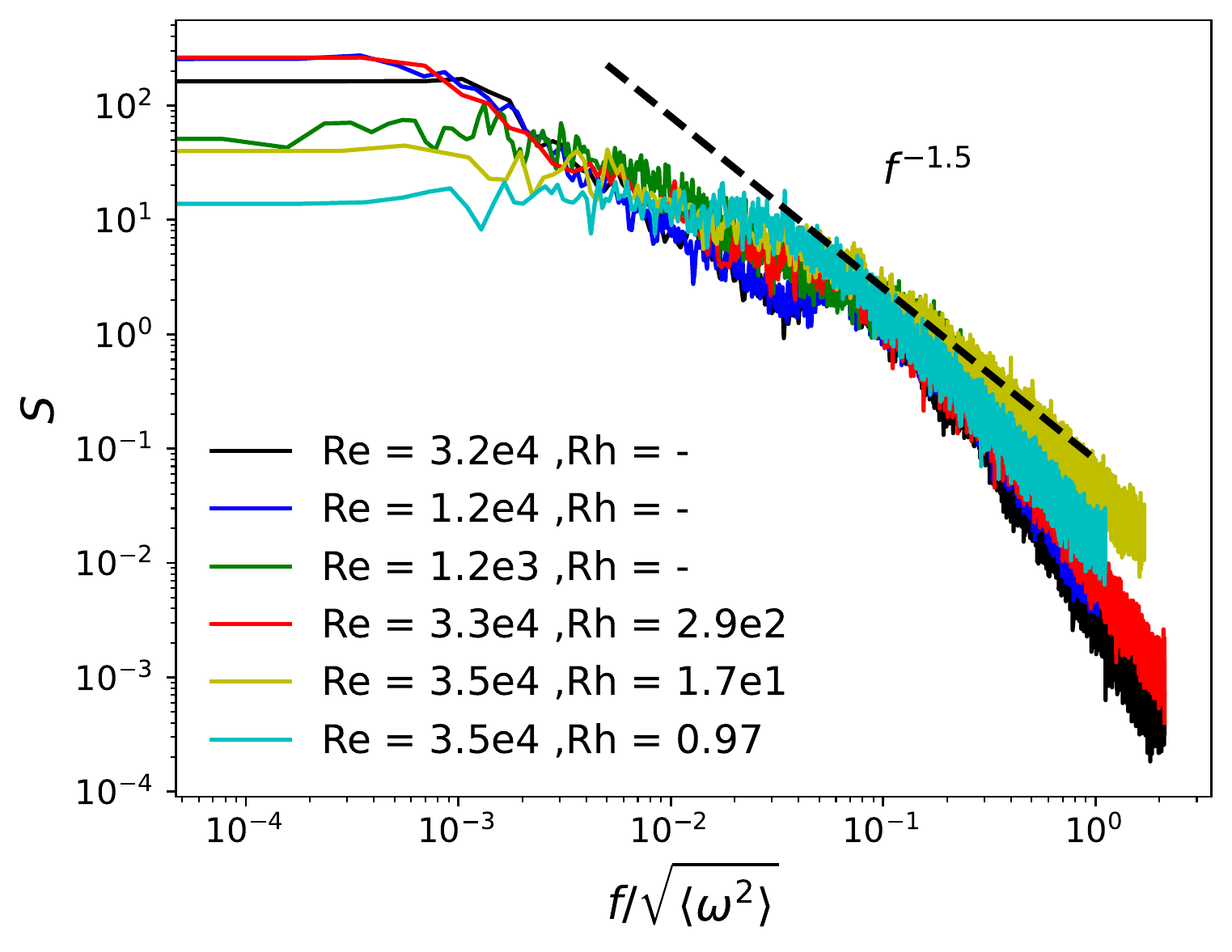}
         \caption{}
         \label{fig:turb_spec}
     \end{subfigure}
     \begin{subfigure}[b]{0.45\textwidth}
         \centering
         \includegraphics[width=\textwidth]{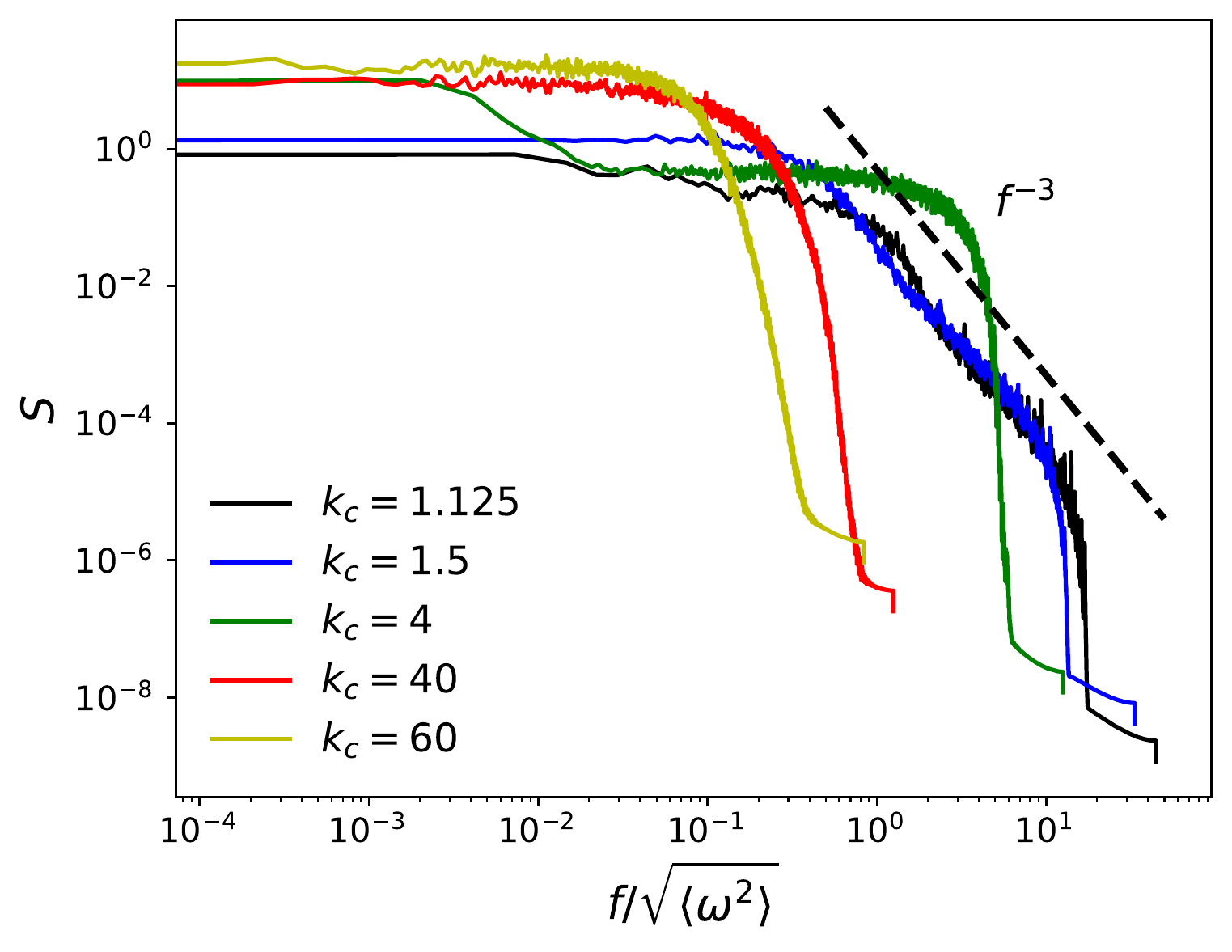}
         \caption{}
         \label{fig:TEE_spec}
     \end{subfigure}
\caption{\label{fig:spec} The figure shows the power spectrum of the $\omega^{\text{min}}$ for a) Turbulence, and b) TEE system for different parameters mentioned in the legend. The dashed lines show the power laws $f^{-1.5}, f^{-3}$ for comparison. }
\end{figure}

To understand the obtained distributions of extreme values, we look at the power spectrum of the time series of the flow variables at a fixed spatial location. Extreme value statistics of time series with power spectrum having $1/f^{\alpha}$ statistics with $0<\alpha<1$ were found to still fall into the FTG class of extreme value statistics. While the variables with steeper decay of power spectrum with $\alpha > 1$ can deviate from the FTG class,  \cite{gyorgyi2007maximal, moloney2011order, leblanc2013universal}. To study the deviation between the distributions obtained from turbulence and the truncated Euler equations, we look at the power spectrum of the time series of vorticity at a single spatial location. Figures \ref{fig:spec} show the power spectrum of the vorticity field at a point for a) the Navier-Stokes turbulence and b) TEE equations for different parameters as mentioned in the legend. The vorticity and the frequency are normalized with the square root of the enstrophy $\sqrt{\left\langle \omega^2 \right\rangle}$. In the case of two-dimensional turbulence the power spectrum decays as a power law with an exponent steeper than $1$. Whereas for the TEE model, for large values of $k_c$ the power spectrum is flat at low frequencies and as we approach order $1$ frequency the spectrum decays exponentially. We see that for large $k_c$ values the power spectrum displays close to a flat spectrum and this weak correlation can lead to the Gumbel statistics for the spatial extremes. For lower values of $k_c$ the spectrum shows a regime of power law behaviour with the exponent being steeper than $1$. The case of turbulence and the low $k_c$ values of the TEE model show strong correlation and their limiting distribution for the extremes deviate from FTG theory. 

\section{Discussion}

Using numerical simulations we have studied the spatial extrema of the vorticity and the determinant of the strain rate tensor for a two-dimensional turbulent flow and the truncated Euler equations. The vorticity extrema are found at the centers of vortices while the extrema of the determinant of the strain rate tensor is found to be at the edges of the vortices. The vorticity distribution at a spatial location is found to be Gaussian in the case of the TEE model, and for the spatial extremes it gives the Gumbel distribution. While for the turbulent flow, the vorticity distribution has power law tails, and leads to a non-trivial distribution for the extremes. The moments of the distribution and its shape are dependent on the non-dimensional parameters $\Rey, \Rh$ studied here, it remains to be seen whether the form of the distribution reaches an asymptote in the limit of zero dissipation coefficients. Using the power spectrum of the flow variables at a spatial location, we find that the TEE model leads to weak temporal correlations in the vorticity field thus leading to the FTG distribution for the spatial extreme distribution. While for the turbulent flow with strong temporal correlations in the vorticity field, the system exhibits a non-trivial distribution for the spatial extreme. 

Studies by \cite{gyorgyi2007maximal, moloney2011order, leblanc2013universal} on signals with a spectrum $f^{-\alpha}$ with $\alpha > 1$ has shown extreme value distributions which do not fall in the FTG class and the limiting distributions in such cases are also sensitive to boundary conditions \cite{majumdar2004exact}. In the turbulent regime due to the dual cascade in two-dimensional turbulence we expect to see multiple power law scalings and future theoretical models can help us determine the distributions for noise that exhibit such correlation. Finally the TEE model leads to a distribution which seems to fall under the FTG class of distributions for large values of $k_c$, for smaller values of $k_c$ it is not clear whether the distributions fall into the FTG class of distributions. One can borrow ideas from the microcanonical ensemble theory to derive the exact distributions for the TEE model. 


\begin{acknowledgments}
We wish to thank the computing resources and support provided by PARAM Shakti supercomputing facility of IIT Kharagpur established under National Supercomputing Mission (NSM), Government of India and supported by Centre for Development of Advanced Computing (CDAC), Pune. We acknowledge support from NSM Grant No. DST/NSM/R\&D\_HPC\_Applications/2021/03.11, from the Institute Scheme from Innovative Research and Development (ISIRD), IIT Kharagpur, Grant No. IIT/SRIC/ISIRD/2021–2022/08 and the Start-up Research Grant No. SRG/2021/001229 from Science \& Engineering Research Board (SERB), India. We thank Alexandros Alexakis for providing useful comments to improve the article. 
\end{acknowledgments}

\appendix

\section{Numerical convergence} \label{App:app_1}

\begin{figure*}
     \centering
     \begin{subfigure}[b]{0.329\textwidth}
         \centering
         \includegraphics[width=\textwidth]{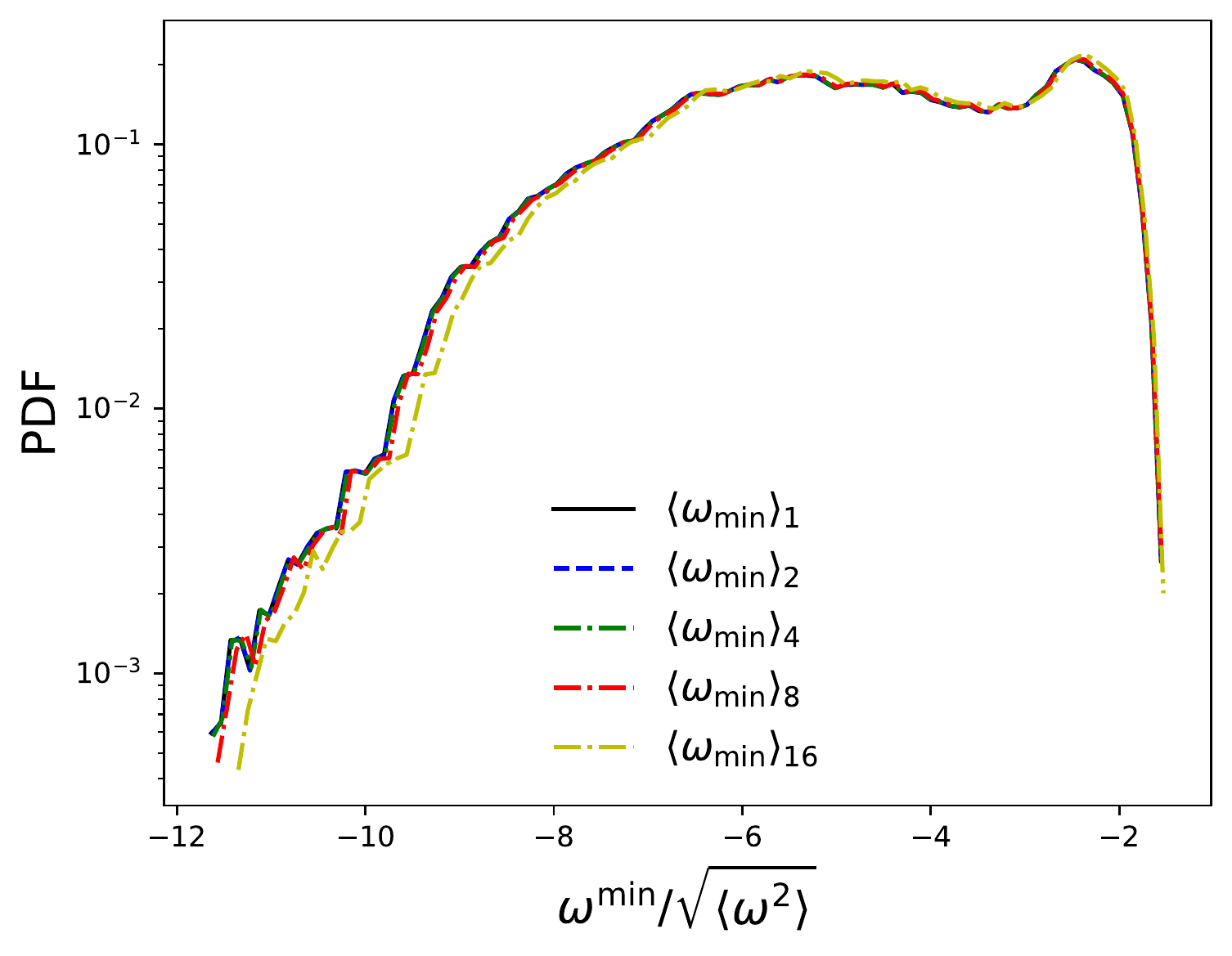}
         \caption{}
         \label{fig:three_graphs_app_1}
     \end{subfigure}
     \hfill
     \begin{subfigure}[b]{0.329\textwidth}
         \centering
         \includegraphics[width=\textwidth]{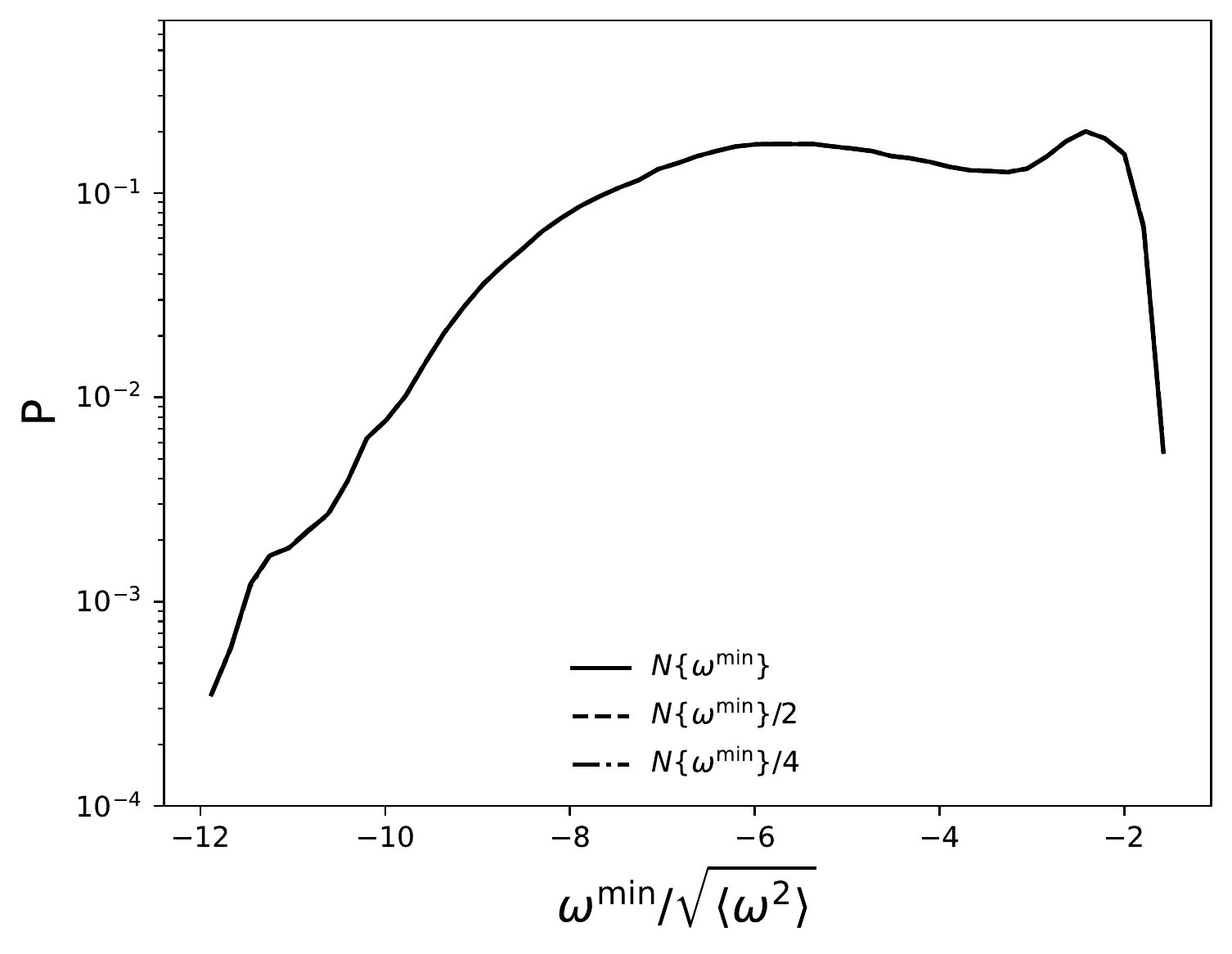}
         \caption{}
         \label{fig:samp}
     \end{subfigure}
     \hfill
     \begin{subfigure}[b]{0.329\textwidth}
         \centering
         \includegraphics[width=\textwidth]{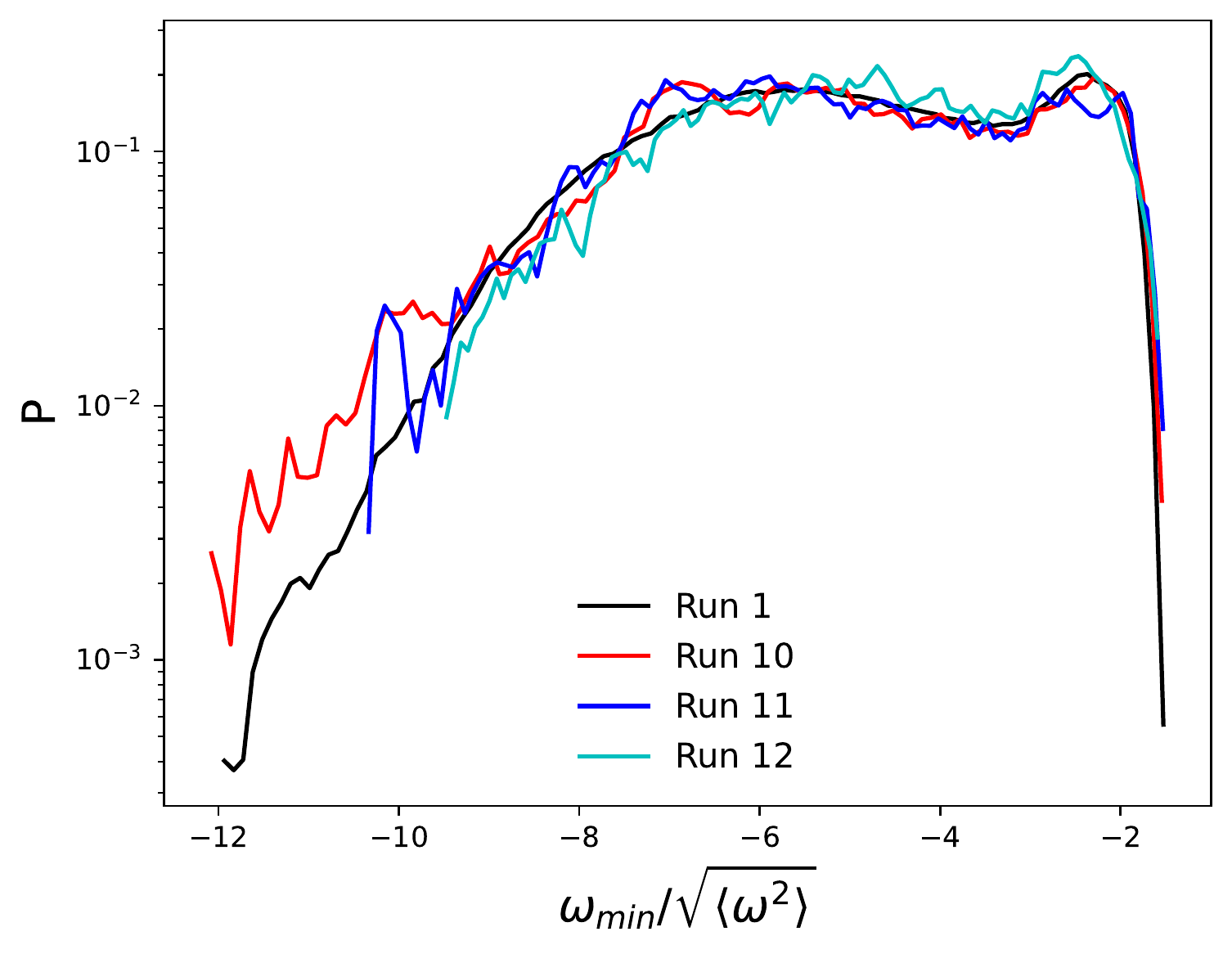}
         \caption{}
         \label{fig:resol}
     \end{subfigure}
        \caption{Figures show the normalised distributions of the minimum of the vorticity for three different series of runs. Figure a) shows the minimum of the locally average fields $\omega^{\text{min}}$. Figure b) shows the distributions when the sampling rate is changed. Figure c) shows the distributions when the resolutions and time steps are changed. All the simulations are done at $\Rey = 3.5 \times 10^4$ in the absence of large scale friction. }
        \label{fig:checks}
\end{figure*}

The spatial extrema occurs at a grid point in the domain of the system, hence an under resolved simulation might have an averaged effect from the other grid points near the extrema. We quantify the extrema obtained from locally averaged fields and see their deviation from the profiles which are not locally averaged. For a given function $f(x, y, t)$, we define the locally averaged function as $\left\langle f(x,y,t) \right\rangle_n$ where the subscript $n$ denotes averaging over $2n-1$ number of nearby points along each direction, taking into account the square geometry this leads to an averaging over $(2n - 1)^2$ points. The extremum is then taken over this local spatially averaged field which is denoted by $\left\langle f(x,y,t) \right\rangle_n^{\text{min}}$ for the function $f(x,y,t)$. Figure \ref{fig:three_graphs_app_1} shows the distribution for locally averaged spatial extrema of the vorticity field $\omega^{\text{min}}$. The comparison between the averages for $n = 1, 2, 4, 8, 16$ show that the distribution is not affected by local averaging over a small domain of points ($n = 1, 2, 4$), for larger number of points (for $n.= 8, 16$) we see that the distribution is slightly affected at the tails . 

Recent studies have also mentioned the importance of having sufficient spatial and temporal resolution in order to capture the distribution of extreme value statistics, \cite{yeung2018effects}. We show in Fig \ref{fig:samp} the distribution of $\omega^{\text{min}}$ for different sampling rates where $N\left\{ \cdot \right\}$ denotes the total number of statistics taken over the whole duration of the simulation. We compare three sets, one with the original sampling, the other two are at half and one fourth sampling rate. We observe that the distribution is not affected by the sampling frequency. We also compare the simulations with larger and smaller resolution with different CFL conditions, these are Runs 10, 11 \& 12 whose parameters are mentioned in Table \ref{tab:table1}. Figure \ref{fig:resol} shows the distributions of $\omega^{\text{min}}$ for the Runs 1, 10, 11 and 12 . The distributions fall on top of each other for all the different simulations with varied spatial and temporal resolutions. In this section we have shown that for the maximum $\Rey$ that has been studied, the distributions converge and they do not depend on the discretisation or sampling schemes. This has been verified for other simulations shown in Table \ref{tab:table1}.


\bibliography{aipsamp}

\providecommand{\noopsort}[1]{}\providecommand{\singleletter}[1]{#1}%
\begin{thebibliography}{38}%
\makeatletter
\providecommand \@ifxundefined [1]{%
 \@ifx{#1\undefined}
}%
\providecommand \@ifnum [1]{%
 \ifnum #1\expandafter \@firstoftwo
 \else \expandafter \@secondoftwo
 \fi
}%
\providecommand \@ifx [1]{%
 \ifx #1\expandafter \@firstoftwo
 \else \expandafter \@secondoftwo
 \fi
}%
\providecommand \natexlab [1]{#1}%
\providecommand \enquote  [1]{``#1''}%
\providecommand \bibnamefont  [1]{#1}%
\providecommand \bibfnamefont [1]{#1}%
\providecommand \citenamefont [1]{#1}%
\providecommand \href@noop [0]{\@secondoftwo}%
\providecommand \href [0]{\begingroup \@sanitize@url \@href}%
\providecommand \@href[1]{\@@startlink{#1}\@@href}%
\providecommand \@@href[1]{\endgroup#1\@@endlink}%
\providecommand \@sanitize@url [0]{\catcode `\\12\catcode `\$12\catcode
  `\&12\catcode `\#12\catcode `\^12\catcode `\_12\catcode `\%12\relax}%
\providecommand \@@startlink[1]{}%
\providecommand \@@endlink[0]{}%
\providecommand \url  [0]{\begingroup\@sanitize@url \@url }%
\providecommand \@url [1]{\endgroup\@href {#1}{\urlprefix }}%
\providecommand \urlprefix  [0]{URL }%
\providecommand \Eprint [0]{\href }%
\providecommand \doibase [0]{http://dx.doi.org/}%
\providecommand \selectlanguage [0]{\@gobble}%
\providecommand \bibinfo  [0]{\@secondoftwo}%
\providecommand \bibfield  [0]{\@secondoftwo}%
\providecommand \translation [1]{[#1]}%
\providecommand \BibitemOpen [0]{}%
\providecommand \bibitemStop [0]{}%
\providecommand \bibitemNoStop [0]{.\EOS\space}%
\providecommand \EOS [0]{\spacefactor3000\relax}%
\providecommand \BibitemShut  [1]{\csname bibitem#1\endcsname}%
\let\auto@bib@innerbib\@empty
\bibitem [{\citenamefont {Yeung}\ \emph {et~al.}(2015)\citenamefont {Yeung},
  \citenamefont {Zhai},\ and\ \citenamefont {Sreenivasan}}]{yeung2015extreme}%
  \BibitemOpen
  \bibfield  {author} {\bibinfo {author} {\bibfnamefont {P.}~\bibnamefont
  {Yeung}}, \bibinfo {author} {\bibfnamefont {X.}~\bibnamefont {Zhai}}, \ and\
  \bibinfo {author} {\bibfnamefont {K.~R.}\ \bibnamefont {Sreenivasan}},\
  }\href@noop {} {\bibfield  {journal} {\bibinfo  {journal} {Proceedings of the
  National Academy of Sciences}\ }\textbf {\bibinfo {volume} {112}},\ \bibinfo
  {pages} {12633} (\bibinfo {year} {2015})}\BibitemShut {NoStop}%
\bibitem [{\citenamefont {Buaria}\ \emph {et~al.}(2019)\citenamefont {Buaria},
  \citenamefont {Pumir}, \citenamefont {Bodenschatz},\ and\ \citenamefont
  {Yeung}}]{buaria2019extreme}%
  \BibitemOpen
  \bibfield  {author} {\bibinfo {author} {\bibfnamefont {D.}~\bibnamefont
  {Buaria}}, \bibinfo {author} {\bibfnamefont {A.}~\bibnamefont {Pumir}},
  \bibinfo {author} {\bibfnamefont {E.}~\bibnamefont {Bodenschatz}}, \ and\
  \bibinfo {author} {\bibfnamefont {P.-K.}\ \bibnamefont {Yeung}},\ }\href@noop
  {} {\bibfield  {journal} {\bibinfo  {journal} {New Journal of Physics}\
  }\textbf {\bibinfo {volume} {21}},\ \bibinfo {pages} {043004} (\bibinfo
  {year} {2019})}\BibitemShut {NoStop}%
\bibitem [{\citenamefont {Yeung}\ and\ \citenamefont
  {Ravikumar}(2020)}]{yeung2020advancing}%
  \BibitemOpen
  \bibfield  {author} {\bibinfo {author} {\bibfnamefont {P.}~\bibnamefont
  {Yeung}}\ and\ \bibinfo {author} {\bibfnamefont {K.}~\bibnamefont
  {Ravikumar}},\ }\href@noop {} {\bibfield  {journal} {\bibinfo  {journal}
  {Physical Review Fluids}\ }\textbf {\bibinfo {volume} {5}},\ \bibinfo {pages}
  {110517} (\bibinfo {year} {2020})}\BibitemShut {NoStop}%
\bibitem [{\citenamefont {Easterling}\ \emph {et~al.}(2000)\citenamefont
  {Easterling}, \citenamefont {Meehl}, \citenamefont {Parmesan}, \citenamefont
  {Changnon}, \citenamefont {Karl},\ and\ \citenamefont
  {Mearns}}]{easterling2000climate}%
  \BibitemOpen
  \bibfield  {author} {\bibinfo {author} {\bibfnamefont {D.~R.}\ \bibnamefont
  {Easterling}}, \bibinfo {author} {\bibfnamefont {G.~A.}\ \bibnamefont
  {Meehl}}, \bibinfo {author} {\bibfnamefont {C.}~\bibnamefont {Parmesan}},
  \bibinfo {author} {\bibfnamefont {S.~A.}\ \bibnamefont {Changnon}}, \bibinfo
  {author} {\bibfnamefont {T.~R.}\ \bibnamefont {Karl}}, \ and\ \bibinfo
  {author} {\bibfnamefont {L.~O.}\ \bibnamefont {Mearns}},\ }\href@noop {}
  {\bibfield  {journal} {\bibinfo  {journal} {science}\ }\textbf {\bibinfo
  {volume} {289}},\ \bibinfo {pages} {2068} (\bibinfo {year}
  {2000})}\BibitemShut {NoStop}%
\bibitem [{\citenamefont {Ragone}\ \emph {et~al.}(2018)\citenamefont {Ragone},
  \citenamefont {Wouters},\ and\ \citenamefont
  {Bouchet}}]{ragone2018computation}%
  \BibitemOpen
  \bibfield  {author} {\bibinfo {author} {\bibfnamefont {F.}~\bibnamefont
  {Ragone}}, \bibinfo {author} {\bibfnamefont {J.}~\bibnamefont {Wouters}}, \
  and\ \bibinfo {author} {\bibfnamefont {F.}~\bibnamefont {Bouchet}},\
  }\href@noop {} {\bibfield  {journal} {\bibinfo  {journal} {Proceedings of the
  National Academy of Sciences}\ }\textbf {\bibinfo {volume} {115}},\ \bibinfo
  {pages} {24} (\bibinfo {year} {2018})}\BibitemShut {NoStop}%
\bibitem [{\citenamefont {Dysthe}\ \emph {et~al.}(2008)\citenamefont {Dysthe},
  \citenamefont {Krogstad},\ and\ \citenamefont
  {M{\"u}ller}}]{dysthe2008oceanic}%
  \BibitemOpen
  \bibfield  {author} {\bibinfo {author} {\bibfnamefont {K.}~\bibnamefont
  {Dysthe}}, \bibinfo {author} {\bibfnamefont {H.~E.}\ \bibnamefont
  {Krogstad}}, \ and\ \bibinfo {author} {\bibfnamefont {P.}~\bibnamefont
  {M{\"u}ller}},\ }\href@noop {} {\bibfield  {journal} {\bibinfo  {journal}
  {Annu. Rev. Fluid Mech.}\ }\textbf {\bibinfo {volume} {40}},\ \bibinfo
  {pages} {287} (\bibinfo {year} {2008})}\BibitemShut {NoStop}%
\bibitem [{\citenamefont {Onorato}\ \emph {et~al.}(2013)\citenamefont
  {Onorato}, \citenamefont {Residori}, \citenamefont {Bortolozzo},
  \citenamefont {Montina},\ and\ \citenamefont {Arecchi}}]{onorato2013rogue}%
  \BibitemOpen
  \bibfield  {author} {\bibinfo {author} {\bibfnamefont {M.}~\bibnamefont
  {Onorato}}, \bibinfo {author} {\bibfnamefont {S.}~\bibnamefont {Residori}},
  \bibinfo {author} {\bibfnamefont {U.}~\bibnamefont {Bortolozzo}}, \bibinfo
  {author} {\bibfnamefont {A.}~\bibnamefont {Montina}}, \ and\ \bibinfo
  {author} {\bibfnamefont {F.}~\bibnamefont {Arecchi}},\ }\href@noop {}
  {\bibfield  {journal} {\bibinfo  {journal} {Physics Reports}\ }\textbf
  {\bibinfo {volume} {528}},\ \bibinfo {pages} {47} (\bibinfo {year}
  {2013})}\BibitemShut {NoStop}%
\bibitem [{\citenamefont {Sch{\"a}fer}\ \emph {et~al.}(2018)\citenamefont
  {Sch{\"a}fer}, \citenamefont {Beck}, \citenamefont {Aihara}, \citenamefont
  {Witthaut},\ and\ \citenamefont {Timme}}]{grid_fluctuations}%
  \BibitemOpen
  \bibfield  {author} {\bibinfo {author} {\bibfnamefont {B.}~\bibnamefont
  {Sch{\"a}fer}}, \bibinfo {author} {\bibfnamefont {C.}~\bibnamefont {Beck}},
  \bibinfo {author} {\bibfnamefont {K.}~\bibnamefont {Aihara}}, \bibinfo
  {author} {\bibfnamefont {D.}~\bibnamefont {Witthaut}}, \ and\ \bibinfo
  {author} {\bibfnamefont {M.}~\bibnamefont {Timme}},\ }\href {\doibase
  10.1038/s41560-017-0058-z} {\bibfield  {journal} {\bibinfo  {journal} {Nature
  Energy}\ }\textbf {\bibinfo {volume} {3}},\ \bibinfo {pages} {119} (\bibinfo
  {year} {2018})}\BibitemShut {NoStop}%
\bibitem [{\citenamefont {Picardo}\ \emph {et~al.}(2023)\citenamefont
  {Picardo}, \citenamefont {Plan},\ and\ \citenamefont
  {Vincenzi}}]{picardo2023polymers}%
  \BibitemOpen
  \bibfield  {author} {\bibinfo {author} {\bibfnamefont {J.~R.}\ \bibnamefont
  {Picardo}}, \bibinfo {author} {\bibfnamefont {E.~L.}\ \bibnamefont {Plan}}, \
  and\ \bibinfo {author} {\bibfnamefont {D.}~\bibnamefont {Vincenzi}},\
  }\href@noop {} {\bibfield  {journal} {\bibinfo  {journal} {arXiv preprint
  arXiv:2301.02990}\ } (\bibinfo {year} {2023})}\BibitemShut {NoStop}%
\bibitem [{\citenamefont {Farazmand}\ and\ \citenamefont
  {Sapsis}(2019)}]{farazmand2019closed}%
  \BibitemOpen
  \bibfield  {author} {\bibinfo {author} {\bibfnamefont {M.}~\bibnamefont
  {Farazmand}}\ and\ \bibinfo {author} {\bibfnamefont {T.~P.}\ \bibnamefont
  {Sapsis}},\ }\href@noop {} {\bibfield  {journal} {\bibinfo  {journal} {Phys.
  Rev. E}\ }\textbf {\bibinfo {volume} {100}},\ \bibinfo {pages} {033110}
  (\bibinfo {year} {2019})}\BibitemShut {NoStop}%
\bibitem [{\citenamefont {Sapsis}(2021)}]{sapsis2021statistics}%
  \BibitemOpen
  \bibfield  {author} {\bibinfo {author} {\bibfnamefont {T.~P.}\ \bibnamefont
  {Sapsis}},\ }\href@noop {} {\bibfield  {journal} {\bibinfo  {journal} {Annu.
  Rev. Fluid Mech.}\ }\textbf {\bibinfo {volume} {53}},\ \bibinfo {pages} {85}
  (\bibinfo {year} {2021})}\BibitemShut {NoStop}%
\bibitem [{\citenamefont {Antal}\ \emph {et~al.}(2001)\citenamefont {Antal},
  \citenamefont {Droz}, \citenamefont {Gy{\"o}rgyi},\ and\ \citenamefont
  {R{\'a}cz}}]{antal20011}%
  \BibitemOpen
  \bibfield  {author} {\bibinfo {author} {\bibfnamefont {T.}~\bibnamefont
  {Antal}}, \bibinfo {author} {\bibfnamefont {M.}~\bibnamefont {Droz}},
  \bibinfo {author} {\bibfnamefont {G.}~\bibnamefont {Gy{\"o}rgyi}}, \ and\
  \bibinfo {author} {\bibfnamefont {Z.}~\bibnamefont {R{\'a}cz}},\ }\href@noop
  {} {\bibfield  {journal} {\bibinfo  {journal} {Physical review letters}\
  }\textbf {\bibinfo {volume} {87}},\ \bibinfo {pages} {240601} (\bibinfo
  {year} {2001})}\BibitemShut {NoStop}%
\bibitem [{\citenamefont {Majumdar}\ \emph {et~al.}(2020)\citenamefont
  {Majumdar}, \citenamefont {Pal},\ and\ \citenamefont
  {Schehr}}]{majumdar2020extreme}%
  \BibitemOpen
  \bibfield  {author} {\bibinfo {author} {\bibfnamefont {S.~N.}\ \bibnamefont
  {Majumdar}}, \bibinfo {author} {\bibfnamefont {A.}~\bibnamefont {Pal}}, \
  and\ \bibinfo {author} {\bibfnamefont {G.}~\bibnamefont {Schehr}},\
  }\href@noop {} {\bibfield  {journal} {\bibinfo  {journal} {Physics Reports}\
  }\textbf {\bibinfo {volume} {840}},\ \bibinfo {pages} {1} (\bibinfo {year}
  {2020})}\BibitemShut {NoStop}%
\bibitem [{\citenamefont {Raychaudhuri}\ \emph {et~al.}(2001)\citenamefont
  {Raychaudhuri}, \citenamefont {Cranston}, \citenamefont {Przybyla},\ and\
  \citenamefont {Shapir}}]{raychaudhuri2001maximal}%
  \BibitemOpen
  \bibfield  {author} {\bibinfo {author} {\bibfnamefont {S.}~\bibnamefont
  {Raychaudhuri}}, \bibinfo {author} {\bibfnamefont {M.}~\bibnamefont
  {Cranston}}, \bibinfo {author} {\bibfnamefont {C.}~\bibnamefont {Przybyla}},
  \ and\ \bibinfo {author} {\bibfnamefont {Y.}~\bibnamefont {Shapir}},\
  }\href@noop {} {\bibfield  {journal} {\bibinfo  {journal} {Physical review
  letters}\ }\textbf {\bibinfo {volume} {87}},\ \bibinfo {pages} {136101}
  (\bibinfo {year} {2001})}\BibitemShut {NoStop}%
\bibitem [{\citenamefont {Gy{\"o}rgyi}\ \emph {et~al.}(2003)\citenamefont
  {Gy{\"o}rgyi}, \citenamefont {Holdsworth}, \citenamefont {Portelli},\ and\
  \citenamefont {R{\'a}cz}}]{gyorgyi2003statistics}%
  \BibitemOpen
  \bibfield  {author} {\bibinfo {author} {\bibfnamefont {G.}~\bibnamefont
  {Gy{\"o}rgyi}}, \bibinfo {author} {\bibfnamefont {P.}~\bibnamefont
  {Holdsworth}}, \bibinfo {author} {\bibfnamefont {B.}~\bibnamefont
  {Portelli}}, \ and\ \bibinfo {author} {\bibfnamefont {Z.}~\bibnamefont
  {R{\'a}cz}},\ }\href@noop {} {\bibfield  {journal} {\bibinfo  {journal}
  {Physical Review E}\ }\textbf {\bibinfo {volume} {68}},\ \bibinfo {pages}
  {056116} (\bibinfo {year} {2003})}\BibitemShut {NoStop}%
\bibitem [{\citenamefont {Rambeau}\ and\ \citenamefont
  {Schehr}(2010)}]{rambeau2010extremal}%
  \BibitemOpen
  \bibfield  {author} {\bibinfo {author} {\bibfnamefont {J.}~\bibnamefont
  {Rambeau}}\ and\ \bibinfo {author} {\bibfnamefont {G.}~\bibnamefont
  {Schehr}},\ }\href@noop {} {\bibfield  {journal} {\bibinfo  {journal} {EPL
  (Europhysics Letters)}\ }\textbf {\bibinfo {volume} {91}},\ \bibinfo {pages}
  {60006} (\bibinfo {year} {2010})}\BibitemShut {NoStop}%
\bibitem [{\citenamefont {Majumdar}\ and\ \citenamefont
  {Comtet}(2005)}]{majumdar2005airy}%
  \BibitemOpen
  \bibfield  {author} {\bibinfo {author} {\bibfnamefont {S.~N.}\ \bibnamefont
  {Majumdar}}\ and\ \bibinfo {author} {\bibfnamefont {A.}~\bibnamefont
  {Comtet}},\ }\href@noop {} {\bibfield  {journal} {\bibinfo  {journal}
  {Journal of Statistical Physics}\ }\textbf {\bibinfo {volume} {119}},\
  \bibinfo {pages} {777} (\bibinfo {year} {2005})}\BibitemShut {NoStop}%
\bibitem [{\citenamefont {Seshasayanan}\ and\ \citenamefont
  {Gallet}(2020)}]{seshasayanan2020onset}%
  \BibitemOpen
  \bibfield  {author} {\bibinfo {author} {\bibfnamefont {K.}~\bibnamefont
  {Seshasayanan}}\ and\ \bibinfo {author} {\bibfnamefont {B.}~\bibnamefont
  {Gallet}},\ }\href@noop {} {\bibfield  {journal} {\bibinfo  {journal}
  {Journal of Fluid Mechanics}\ }\textbf {\bibinfo {volume} {901}} (\bibinfo
  {year} {2020})}\BibitemShut {NoStop}%
\bibitem [{\citenamefont {Alexakis}\ and\ \citenamefont
  {Biferale}(2018)}]{alexakis2018cascades}%
  \BibitemOpen
  \bibfield  {author} {\bibinfo {author} {\bibfnamefont {A.}~\bibnamefont
  {Alexakis}}\ and\ \bibinfo {author} {\bibfnamefont {L.}~\bibnamefont
  {Biferale}},\ }\href@noop {} {\bibfield  {journal} {\bibinfo  {journal}
  {Physics Reports}\ }\textbf {\bibinfo {volume} {767}},\ \bibinfo {pages} {1}
  (\bibinfo {year} {2018})}\BibitemShut {NoStop}%
\bibitem [{\citenamefont {Seshasayanan}\ \emph {et~al.}(2014)\citenamefont
  {Seshasayanan}, \citenamefont {Benavides},\ and\ \citenamefont
  {Alexakis}}]{seshasayanan2014edge}%
  \BibitemOpen
  \bibfield  {author} {\bibinfo {author} {\bibfnamefont {K.}~\bibnamefont
  {Seshasayanan}}, \bibinfo {author} {\bibfnamefont {S.~J.}\ \bibnamefont
  {Benavides}}, \ and\ \bibinfo {author} {\bibfnamefont {A.}~\bibnamefont
  {Alexakis}},\ }\href@noop {} {\bibfield  {journal} {\bibinfo  {journal}
  {Physical Review E}\ }\textbf {\bibinfo {volume} {90}},\ \bibinfo {pages}
  {051003} (\bibinfo {year} {2014})}\BibitemShut {NoStop}%
\bibitem [{\citenamefont {Benavides}\ and\ \citenamefont
  {Alexakis}(2017)}]{benavides2017critical}%
  \BibitemOpen
  \bibfield  {author} {\bibinfo {author} {\bibfnamefont {S.~J.}\ \bibnamefont
  {Benavides}}\ and\ \bibinfo {author} {\bibfnamefont {A.}~\bibnamefont
  {Alexakis}},\ }\href@noop {} {\bibfield  {journal} {\bibinfo  {journal}
  {Journal of Fluid Mechanics}\ }\textbf {\bibinfo {volume} {822}},\ \bibinfo
  {pages} {364} (\bibinfo {year} {2017})}\BibitemShut {NoStop}%
\bibitem [{\citenamefont {van Kan}\ \emph {et~al.}(2021)\citenamefont {van
  Kan}, \citenamefont {Alexakis},\ and\ \citenamefont
  {Brachet}}]{van2021intermittency}%
  \BibitemOpen
  \bibfield  {author} {\bibinfo {author} {\bibfnamefont {A.}~\bibnamefont {van
  Kan}}, \bibinfo {author} {\bibfnamefont {A.}~\bibnamefont {Alexakis}}, \ and\
  \bibinfo {author} {\bibfnamefont {M.-E.}\ \bibnamefont {Brachet}},\
  }\href@noop {} {\bibfield  {journal} {\bibinfo  {journal} {Physical Review
  E}\ }\textbf {\bibinfo {volume} {103}},\ \bibinfo {pages} {053102} (\bibinfo
  {year} {2021})}\BibitemShut {NoStop}%
\bibitem [{\citenamefont {van Kan}\ and\ \citenamefont
  {P{\'e}tr{\'e}lis}(2023)}]{van20231}%
  \BibitemOpen
  \bibfield  {author} {\bibinfo {author} {\bibfnamefont {A.}~\bibnamefont {van
  Kan}}\ and\ \bibinfo {author} {\bibfnamefont {F.}~\bibnamefont
  {P{\'e}tr{\'e}lis}},\ }\href@noop {} {\bibfield  {journal} {\bibinfo
  {journal} {Journal of Statistical Mechanics: Theory and Experiment}\ }\textbf
  {\bibinfo {volume} {2023}},\ \bibinfo {pages} {013204} (\bibinfo {year}
  {2023})}\BibitemShut {NoStop}%
\bibitem [{\citenamefont {Alexakis}(2015)}]{alexakis2015rotating}%
  \BibitemOpen
  \bibfield  {author} {\bibinfo {author} {\bibfnamefont {A.}~\bibnamefont
  {Alexakis}},\ }\href@noop {} {\bibfield  {journal} {\bibinfo  {journal}
  {Journal of Fluid Mechanics}\ }\textbf {\bibinfo {volume} {769}},\ \bibinfo
  {pages} {46} (\bibinfo {year} {2015})}\BibitemShut {NoStop}%
\bibitem [{\citenamefont {Favier}\ \emph {et~al.}(2019)\citenamefont {Favier},
  \citenamefont {Guervilly},\ and\ \citenamefont
  {Knobloch}}]{favier2019subcritical}%
  \BibitemOpen
  \bibfield  {author} {\bibinfo {author} {\bibfnamefont {B.}~\bibnamefont
  {Favier}}, \bibinfo {author} {\bibfnamefont {C.}~\bibnamefont {Guervilly}}, \
  and\ \bibinfo {author} {\bibfnamefont {E.}~\bibnamefont {Knobloch}},\
  }\href@noop {} {\bibfield  {journal} {\bibinfo  {journal} {Journal of Fluid
  Mechanics}\ }\textbf {\bibinfo {volume} {864}} (\bibinfo {year}
  {2019})}\BibitemShut {NoStop}%
\bibitem [{\citenamefont {Herault}\ \emph {et~al.}(2015)\citenamefont
  {Herault}, \citenamefont {P{\'e}tr{\'e}lis},\ and\ \citenamefont
  {Fauve}}]{herault20151}%
  \BibitemOpen
  \bibfield  {author} {\bibinfo {author} {\bibfnamefont {J.}~\bibnamefont
  {Herault}}, \bibinfo {author} {\bibfnamefont {F.}~\bibnamefont
  {P{\'e}tr{\'e}lis}}, \ and\ \bibinfo {author} {\bibfnamefont
  {S.}~\bibnamefont {Fauve}},\ }\href@noop {} {\bibfield  {journal} {\bibinfo
  {journal} {Journal of Statistical Physics}\ }\textbf {\bibinfo {volume}
  {161}},\ \bibinfo {pages} {1379} (\bibinfo {year} {2015})}\BibitemShut
  {NoStop}%
\bibitem [{\citenamefont {Dallas}\ \emph {et~al.}(2020)\citenamefont {Dallas},
  \citenamefont {Seshasayanan},\ and\ \citenamefont
  {Fauve}}]{dallas2020transitions}%
  \BibitemOpen
  \bibfield  {author} {\bibinfo {author} {\bibfnamefont {V.}~\bibnamefont
  {Dallas}}, \bibinfo {author} {\bibfnamefont {K.}~\bibnamefont
  {Seshasayanan}}, \ and\ \bibinfo {author} {\bibfnamefont {S.}~\bibnamefont
  {Fauve}},\ }\href@noop {} {\bibfield  {journal} {\bibinfo  {journal}
  {Physical Review Fluids}\ }\textbf {\bibinfo {volume} {5}},\ \bibinfo {pages}
  {084610} (\bibinfo {year} {2020})}\BibitemShut {NoStop}%
\bibitem [{\citenamefont {Tsang}(2010)}]{tsang2010nonuniversal}%
  \BibitemOpen
  \bibfield  {author} {\bibinfo {author} {\bibfnamefont {Y.-K.}\ \bibnamefont
  {Tsang}},\ }\href@noop {} {\bibfield  {journal} {\bibinfo  {journal} {Phys.
  Fluids}\ }\textbf {\bibinfo {volume} {22}},\ \bibinfo {pages} {115102}
  (\bibinfo {year} {2010})}\BibitemShut {NoStop}%
\bibitem [{\citenamefont {Ascher}\ \emph {et~al.}(1997)\citenamefont {Ascher},
  \citenamefont {Ruuth},\ and\ \citenamefont {Spiteri}}]{ascher1997implicit}%
  \BibitemOpen
  \bibfield  {author} {\bibinfo {author} {\bibfnamefont {U.~M.}\ \bibnamefont
  {Ascher}}, \bibinfo {author} {\bibfnamefont {S.~J.}\ \bibnamefont {Ruuth}}, \
  and\ \bibinfo {author} {\bibfnamefont {R.~J.}\ \bibnamefont {Spiteri}},\
  }\href@noop {} {\bibfield  {journal} {\bibinfo  {journal} {Applied Numerical
  Mathematics}\ }\textbf {\bibinfo {volume} {25}},\ \bibinfo {pages} {151}
  (\bibinfo {year} {1997})}\BibitemShut {NoStop}%
\bibitem [{\citenamefont {Alexakis}\ and\ \citenamefont
  {Doering}(2006)}]{alexakis2006energy}%
  \BibitemOpen
  \bibfield  {author} {\bibinfo {author} {\bibfnamefont {A.}~\bibnamefont
  {Alexakis}}\ and\ \bibinfo {author} {\bibfnamefont {C.~R.}\ \bibnamefont
  {Doering}},\ }\href@noop {} {\bibfield  {journal} {\bibinfo  {journal}
  {Physics letters A}\ }\textbf {\bibinfo {volume} {359}},\ \bibinfo {pages}
  {652} (\bibinfo {year} {2006})}\BibitemShut {NoStop}%
\bibitem [{\citenamefont {Gallet}(2015)}]{gallet2015exact}%
  \BibitemOpen
  \bibfield  {author} {\bibinfo {author} {\bibfnamefont {B.}~\bibnamefont
  {Gallet}},\ }\href@noop {} {\bibfield  {journal} {\bibinfo  {journal}
  {Journal of Fluid Mechanics}\ }\textbf {\bibinfo {volume} {783}},\ \bibinfo
  {pages} {412} (\bibinfo {year} {2015})}\BibitemShut {NoStop}%
\bibitem [{\citenamefont {Kraichnan}(1975)}]{kraichnan1975statistical}%
  \BibitemOpen
  \bibfield  {author} {\bibinfo {author} {\bibfnamefont {R.~H.}\ \bibnamefont
  {Kraichnan}},\ }\href@noop {} {\bibfield  {journal} {\bibinfo  {journal}
  {Journal of Fluid Mechanics}\ }\textbf {\bibinfo {volume} {67}},\ \bibinfo
  {pages} {155} (\bibinfo {year} {1975})}\BibitemShut {NoStop}%
\bibitem [{\citenamefont {van Kan}\ \emph {et~al.}(2022)\citenamefont {van
  Kan}, \citenamefont {Alexakis},\ and\ \citenamefont
  {Brachet}}]{van2022geometric}%
  \BibitemOpen
  \bibfield  {author} {\bibinfo {author} {\bibfnamefont {A.}~\bibnamefont {van
  Kan}}, \bibinfo {author} {\bibfnamefont {A.}~\bibnamefont {Alexakis}}, \ and\
  \bibinfo {author} {\bibfnamefont {M.}~\bibnamefont {Brachet}},\ }\href@noop
  {} {\bibfield  {journal} {\bibinfo  {journal} {Philosophical Transactions of
  the Royal Society A}\ }\textbf {\bibinfo {volume} {380}},\ \bibinfo {pages}
  {20210049} (\bibinfo {year} {2022})}\BibitemShut {NoStop}%
\bibitem [{\citenamefont {Gy{\"o}rgyi}\ \emph {et~al.}(2007)\citenamefont
  {Gy{\"o}rgyi}, \citenamefont {Moloney}, \citenamefont {Ozog{\'a}ny},\ and\
  \citenamefont {R{\'a}cz}}]{gyorgyi2007maximal}%
  \BibitemOpen
  \bibfield  {author} {\bibinfo {author} {\bibfnamefont {G.}~\bibnamefont
  {Gy{\"o}rgyi}}, \bibinfo {author} {\bibfnamefont {N.}~\bibnamefont
  {Moloney}}, \bibinfo {author} {\bibfnamefont {K.}~\bibnamefont
  {Ozog{\'a}ny}}, \ and\ \bibinfo {author} {\bibfnamefont {Z.}~\bibnamefont
  {R{\'a}cz}},\ }\href@noop {} {\bibfield  {journal} {\bibinfo  {journal}
  {Physical Review E}\ }\textbf {\bibinfo {volume} {75}},\ \bibinfo {pages}
  {021123} (\bibinfo {year} {2007})}\BibitemShut {NoStop}%
\bibitem [{\citenamefont {Moloney}\ \emph {et~al.}(2011)\citenamefont
  {Moloney}, \citenamefont {Ozog{\'a}ny},\ and\ \citenamefont
  {R{\'a}cz}}]{moloney2011order}%
  \BibitemOpen
  \bibfield  {author} {\bibinfo {author} {\bibfnamefont {N.}~\bibnamefont
  {Moloney}}, \bibinfo {author} {\bibfnamefont {K.}~\bibnamefont
  {Ozog{\'a}ny}}, \ and\ \bibinfo {author} {\bibfnamefont {Z.}~\bibnamefont
  {R{\'a}cz}},\ }\href@noop {} {\bibfield  {journal} {\bibinfo  {journal}
  {Physical Review E}\ }\textbf {\bibinfo {volume} {84}},\ \bibinfo {pages}
  {061101} (\bibinfo {year} {2011})}\BibitemShut {NoStop}%
\bibitem [{\citenamefont {LeBlanc}\ \emph {et~al.}(2013)\citenamefont
  {LeBlanc}, \citenamefont {Angheluta}, \citenamefont {Dahmen},\ and\
  \citenamefont {Goldenfeld}}]{leblanc2013universal}%
  \BibitemOpen
  \bibfield  {author} {\bibinfo {author} {\bibfnamefont {M.}~\bibnamefont
  {LeBlanc}}, \bibinfo {author} {\bibfnamefont {L.}~\bibnamefont {Angheluta}},
  \bibinfo {author} {\bibfnamefont {K.}~\bibnamefont {Dahmen}}, \ and\ \bibinfo
  {author} {\bibfnamefont {N.}~\bibnamefont {Goldenfeld}},\ }\href@noop {}
  {\bibfield  {journal} {\bibinfo  {journal} {Physical Review E}\ }\textbf
  {\bibinfo {volume} {87}},\ \bibinfo {pages} {022126} (\bibinfo {year}
  {2013})}\BibitemShut {NoStop}%
\bibitem [{\citenamefont {Majumdar}\ and\ \citenamefont
  {Comtet}(2004)}]{majumdar2004exact}%
  \BibitemOpen
  \bibfield  {author} {\bibinfo {author} {\bibfnamefont {S.~N.}\ \bibnamefont
  {Majumdar}}\ and\ \bibinfo {author} {\bibfnamefont {A.}~\bibnamefont
  {Comtet}},\ }\href@noop {} {\bibfield  {journal} {\bibinfo  {journal}
  {Physical review letters}\ }\textbf {\bibinfo {volume} {92}},\ \bibinfo
  {pages} {225501} (\bibinfo {year} {2004})}\BibitemShut {NoStop}%
\bibitem [{\citenamefont {Yeung}\ \emph {et~al.}(2018)\citenamefont {Yeung},
  \citenamefont {Sreenivasan},\ and\ \citenamefont {Pope}}]{yeung2018effects}%
  \BibitemOpen
  \bibfield  {author} {\bibinfo {author} {\bibfnamefont {P.}~\bibnamefont
  {Yeung}}, \bibinfo {author} {\bibfnamefont {K.}~\bibnamefont {Sreenivasan}},
  \ and\ \bibinfo {author} {\bibfnamefont {S.}~\bibnamefont {Pope}},\
  }\href@noop {} {\bibfield  {journal} {\bibinfo  {journal} {Physical Review
  Fluids}\ }\textbf {\bibinfo {volume} {3}},\ \bibinfo {pages} {064603}
  (\bibinfo {year} {2018})}\BibitemShut {NoStop}%
\end{thebibliography}%

\end{document}